\documentclass[a4paper,fleqn,usenatbib]{mnras}

\usepackage{newtxtext,newtxmath}

\usepackage[T1]{fontenc}
\usepackage{ae,aecompl}

\usepackage{graphicx}	

\newcommand{\spose}[1]{\hbox  to 0pt{#1\hss}}  
\newcommand{\lta}{\mathrel{\spose{\lower 3pt\hbox{$\sim$}}\raise  2.0pt\hbox{$<$}}}
\newcommand{\gta}{\mathrel{\spose{\lower  3pt\hbox{$\sim$}}\raise 2.0pt\hbox{$>$}}}
\newcommand \hi {\ifmmode \rm {\sc Hi} \else {\sc Hi}\fi}

\newcommand{\ms} {\ifmmode  \,{\rm m\,s}^{-2} \else $\,\rm m\,s^{-2}  $ \fi }
\newcommand{\kms} {\ifmmode  \,{\rm km\,s}^{-1} \else $\,\rm km\,s^{-1}  $ \fi }
\newcommand{\kpc} {\ifmmode  {\rm kpc}  \else ${\rm  kpc}$ \fi  }  
\newcommand{\Msun} {\ifmmode \rm M_{\odot} \else $\rm M_{\odot}$ \fi}

\newcommand{\LCDM} {\ifmmode \Lambda{\rm CDM} \else $\Lambda{\rm CDM}$ \fi}
\newcommand{\Omegam} {\ifmmode \Omega_{\rm m} \else $\Omega_{\rm m}$ \fi} 
\newcommand{\Omegab} {\ifmmode \Omega_{\rm b} \else $\Omega_{\rm b}$ \fi} 
\newcommand{\OmegaL} {\ifmmode \Omega_{\rm \Lambda} \else $\Omega_{\rm \Lambda}$\fi} 
\newcommand{\rhocrit} {\ifmmode \rho_{\rm crit} \else $\rho_{\rm crit}$ \fi}

\newcommand{\Vmaxdmo} {\ifmmode V_{\rm  max}^{\rm DMO} \else  $V_{\rm max}^{\rm DMO}$  \fi} 
\newcommand{\Mhalo} {\ifmmode M_{\rm halo} \else $M_{\rm  halo}$ \fi}  

\newcommand{\Vflat} {\ifmmode V_{\rm flat} \else $V_{\rm flat}$ \fi}
\newcommand{\VHI} {\ifmmode V_{\rm HI} \else $V_{\rm HI}$ \fi} 
\newcommand{\Mstar} {\ifmmode M_{\rm star} \else $M_{\rm star}$ \fi} 
\newcommand{\Mbar} {\ifmmode M_{\rm bar} \else $M_{\rm bar}$ \fi} 

\newcommand{\alphamin} {\ifmmode \alpha_{\rm min} \else $\alpha_{\rm min}$ \fi}
\newcommand{\abar} {\ifmmode a_{\rm bar} \else $a_{\rm bar}$ \fi}
\newcommand{\adm} {\ifmmode a_{\rm dm} \else $a_{\rm dm}$ \fi}

\title[Origin of MOND phenomenology in LCDM] {NIHAO XVIII: Origin of the MOND phenomenology of galactic rotation curves in a \LCDM universe}

\author[Dutton et al.]{Aaron  A. Dutton,$^{1}$\thanks{dutton@nyu.edu} Andrea V. Macci\`o,$^{1,2}$ Aura Obreja,$^{3,1}$ Tobias Buck$^2$\\
$^1$New York University Abu Dhabi, PO Box 129188, Abu Dhabi, United Arab Emirates\\
$^2$Max-Planck-Institut f\"ur Astronomie, K\"onigstuhl 17, 69117 Heidelberg, Germany\\
$^3$University Observatory Munich, Scheinerstra\ss e 1, D-81679 Munich, Germany\\
}

\pubyear{2019}

\begin{document}
\label{firstpage}
\pagerange{\pageref{firstpage}--\pageref{lastpage}}
\maketitle

\begin{abstract}
{ The phenomenological basis for Modified Newtonian Dynamics (MOND) is
  the radial-acceleration-relation (RAR) between the observed
  acceleration, $a=V^2_{\rm rot}(r)/ r$, and the acceleration
  accounted for by the observed baryons (stars and cold gas), $a_{\rm
    bar}=V_{\rm bar}^2(r)/r$.  We show that the RAR arises naturally
  in the NIHAO sample of 89 high-resolution \LCDM cosmological galaxy
  formation simulations.  The overall scatter from NIHAO is just 0.079
  dex, consistent with observational constraints.  However, we show
  that the scatter depends on stellar mass.  At high masses ($10^9
  \lta \Mstar \lta 10^{11}\Msun$) the simulated scatter is just
  $\simeq 0.04$ dex, increasing to $\simeq 0.11$ dex at low masses
  ($10^7 \lta \Mstar \lta 10^{9}\Msun$). Observations show a similar
  dependence for the intrinsic scatter.  At high masses the intrinsic
  scatter is consistent with the zero scatter assumed by MOND, but at
  low masses the intrinsic scatter is non-zero, strongly disfavoring
  MOND.  Applying MOND to our simulations yields remarkably good fits
  to most of the circular velocity profiles. In cases of mild
  disagreement the stellar mass-to-light ratio and/or ``distance'' can
  be tuned to yield acceptable fits, as is often done in observational
  mass models.  In dwarf galaxies with $\Mstar\sim10^6\Msun$ MOND
  breaks down, predicting lower accelerations than observed and in our
  \LCDM simulations.  The assumptions that MOND is based on (e.g.,
  asymptotically flat rotation curves, zero intrinsic scatter in
    the RAR) are only approximately true in $\LCDM$.  Thus if one
  wishes to go beyond Newtonian dynamics there is more freedom in the
  observed RAR than assumed by MOND.  } 
\end{abstract}


\setcounter{footnote}{1}


\begin{figure*}
  \includegraphics[width=0.45\textwidth]{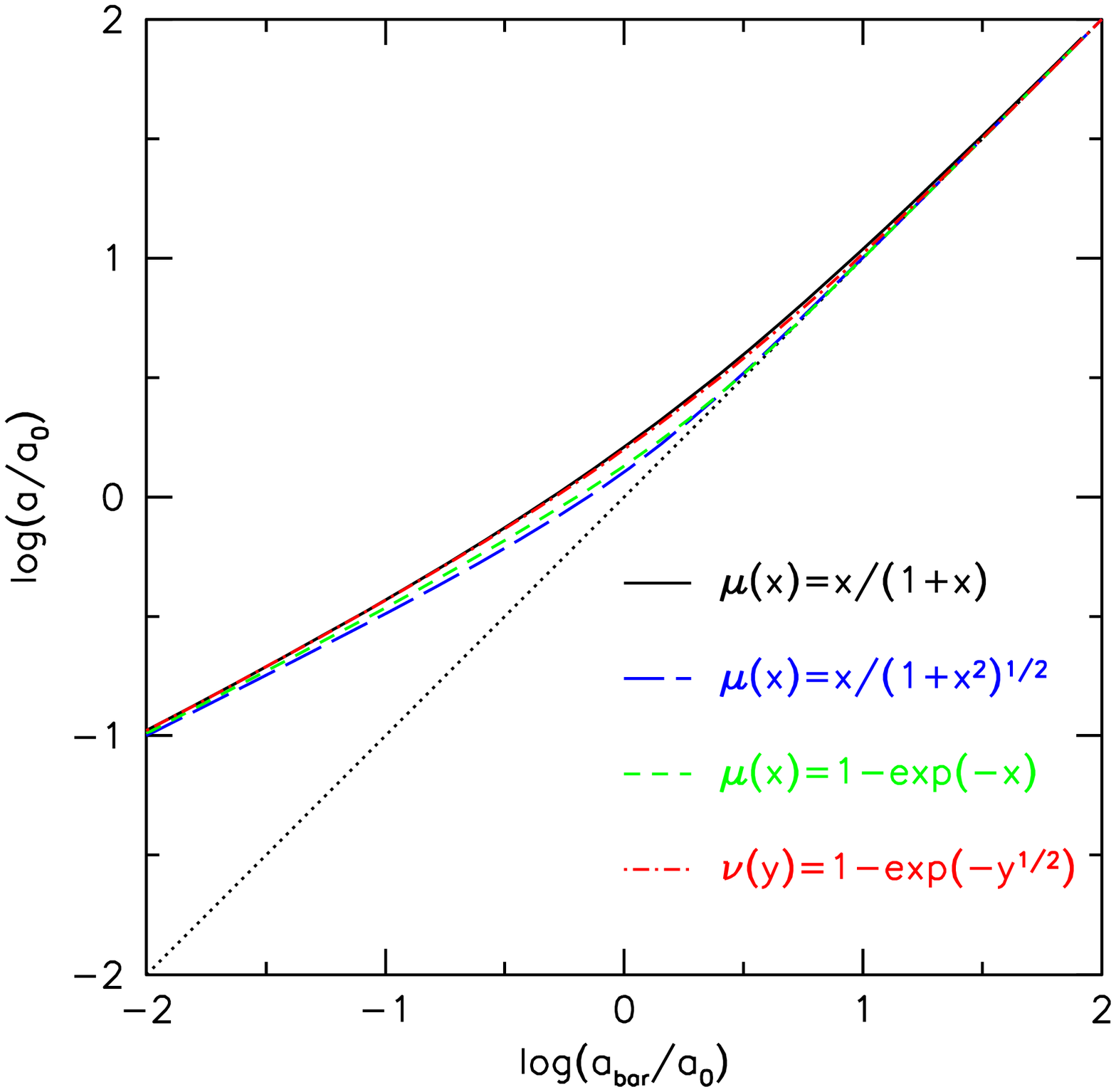}
  \includegraphics[width=0.45\textwidth]{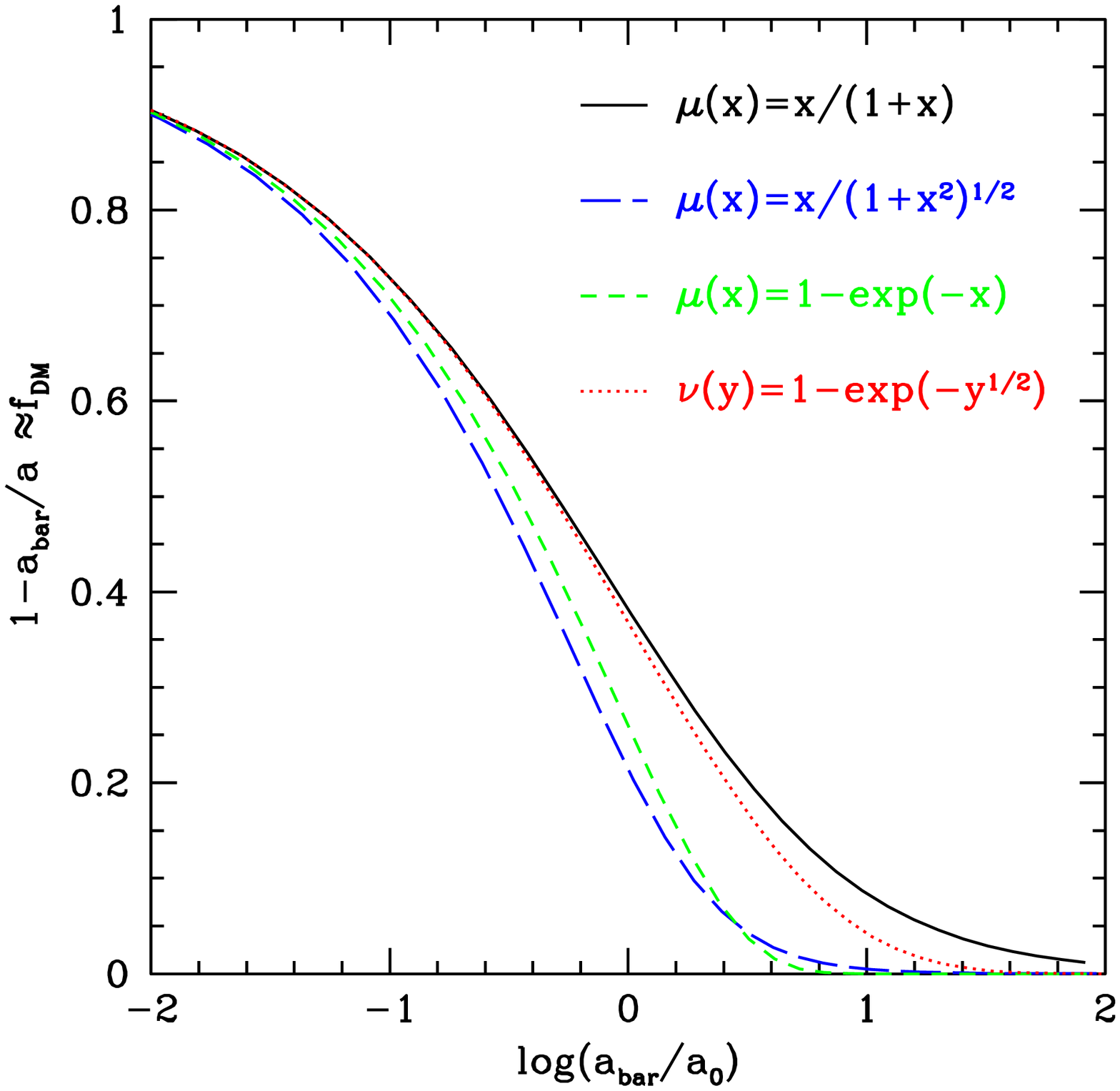}
  \caption{Commonly used MOND interpolation functions of the form
    $a\, \mu(a/a_0)=\abar$, and $a=\abar/\nu(\abar/a_0)$. Left: $a$ vs
    $\abar$ on a log-scale (radial acceleration relation), right:
    $1-\abar/a$ vs $\abar$ ( mass discrepancy - acceleration relation) with
    a linear vertical scale to highlight the differences between
    fitting functions. The dotted line shows the 1:1 relation.}
\label{fig:interp}
\end{figure*}

\section{Introduction}
\label{sec:intro}

The observational evidence for dark matter goes back many
decades. \citet{Zwicky33} showed that if the Virgo cluster is bound,
its total mass must greatly exceed the sum of the masses of the
individual members. In the 1970's galaxy rotation curves were found to
level off at large radii, implying the presence of substantial mass
outside the optically visible dimensions of galaxies \citep{Faber79}.

Further evidence for missing mass comes from the growth of cosmic
structure.  The Cosmic Microwave Background (CMB) is remarkably
uniform with temperature fluctuations $\Delta T/T = \Delta \rho/\rho
\sim 10^{-5}$ \citep{Smoot92}. In order for these density fluctuations
to grow into the observed galaxies $\Delta\rho/\rho\sim 10^6$, they
need to increase by a factor of $10^{11}$!  In the baryon-dominated
universe, in order to have nonlinear structure now, the induced
fluctuations in the CMB would have to be much larger than is observed.  In
a dark matter dominated universe, dark matter perturbations grow prior
to recombination, and this induces a rapid growth in baryon
perturbations shortly after recombination, allowing nonlinear
structure to form now from a smaller baryon fluctuation at
recombination \citep{Bond80}.

An alternative to dark matter was proposed by \citet{Milgrom83} who
considered the possibility that rather than there being missing mass
in galaxies, there is modification to Newtonian dynamics (MOND) at low
accelerations $a<a_0$, where $a_0\sim 10^{-10}$ m s$^{-2}$.
The observed centripetal acceleration (in the
plane of the disk) is defined as
\begin{equation}
  a(r)=V^2(r)/r=-\partial \Phi(r) / \partial r,
\end{equation}
where $\Phi$ is the total gravitational potential.  Likewise, the
acceleration predicted due to the observed baryons, $a_{\rm bar} =
V_{\rm bar}^2 / r$. The following assumptions restrict the form of the
modification to Newtonian dynamics.

{\bf Assumption 1:} there is no missing mass at high accelerations. Thus
\begin{equation}
  a(r) = a_{\rm bar}(r), \; (a \gg a_0).
\end{equation}

{\bf Assumption 2:} rotation curves are asymptotically flat at low
accelerations. At large radii the total acceleration is given by $a(r)
= V_{\rm flat}^2 / r$, and the enclosed baryonic mass is a constant:
$a_{\rm bar}(r) = V_{\rm bar}^2(r)/r = G M_{\rm bar} / r^2$. Requiring
that $a$ is a function of $a_{\rm bar}$ and not radius implies
\begin{equation}
  \label{eq:btf}
  V_{\rm flat}^4 = G M_{\rm bar} a^2 / a_{\rm bar}.
\end{equation}
Since $V_{\rm flat}, G$, and $M_{\rm bar}$ are constants, this implies
\begin{equation}
  a^2(r) / a_{\rm bar}(r) \equiv a_0,\;   (a \ll a_0)
\end{equation}
is also a constant. Note that Eq.~\ref{eq:btf} is a Baryonic
Tully-Fisher relation with slope of 4.

{\bf Assumption 3:} there is a unique interpolation function, $\mu(x)$, between the two acceleration extremes:
\begin{equation}
  a(r)\mu(a/a_0) = a_{\rm bar}(r).
\end{equation}
However, this interpolation function is not specified by the theory. A popular choice is
\begin{equation}
  \mu(x) = x / \sqrt{1+x^2}.
\end{equation}
This can be inverted to express $a$ in terms of $a_{\rm bar}$ \citep{vandenBosch00}:
\begin{equation}
  a(r)= a_{\rm bar}(r) \left(\frac{1}{2} +\frac{1}{2}\sqrt{1+4a^2_0/a^2_{\rm bar}}\right)^{1/2}.
\end{equation}
Other choices and their inversions include
\begin{equation}
  \mu(x) = x / (1+x);
\end{equation}
\begin{equation}
  a(r)= a_{\rm bar}(r) \left(\frac{1}{2} +\frac{1}{2}\sqrt{1+4a_0/a_{\rm bar}}\right),
\end{equation}
and
\begin{equation}
  \mu(x) = 1 - \exp(-x);
\end{equation}

Alternatively \citet{McGaugh16} adopt
\begin{equation}
  a(r) = a_{\rm bar}(r) / \nu(a_{\rm bar}/a_0) 
\end{equation}
with
\begin{equation}
\nu(y) = ( 1 - \exp(-\sqrt{y}))
\end{equation}
and find $a_0=1.20\pm0.02$(ran.)$\pm 0.24$(sys.)$\times 10^{-10}\ms$.

Fig.~\ref{fig:interp} shows these interpolation functions. On a
logarithmic scale over 4 orders of magnitude the differences appear
small (left panel). Upon closer inspection the differences are
significant (right panel). At $\abar=a_0$ the mass discrepancy 
  expressed as the dark matter fraction varies from $0.2 \lta f_{\rm
  DM} \lta 0.4$.  At $\abar=10a_0$, the mass discrepancy varies from $0.0
\lta f_{\rm DM} \lta 0.1$. These differences highlight the
non-uniqueness of predictions from MOND.  The relation between total
acceleration and baryonic acceleration (left panel) is often referred
to as the radial acceleration relation (RAR), while the relation
between the mass discrepancy and baryonic acceleration (right panel)
is often referred to as the mass discrepancy acceleration relation
\citep[MDAR,][]{Sanders90, McGaugh04}.

There have been numerous theoretical studies of the RAR in a \LCDM
context, using both analytic models and cosmological hydrodynamical
simulations \citep[e.g.,][]{vandenBosch00, DiCintio16, Desmond17,
  Keller17, Ludlow17, Navarro17}.  These authors have shown that
galaxy formation in a \LCDM universe results in a RAR close to, but
not identically equal to that commonly ascribed to MOND. 

In this paper we use the NIHAO suite of cosmological galaxy formation
simulations \citep{Wang15} to gain insight into the origin of the RAR
in a \LCDM universe. Compared to previous cosmological hydrodynamical
simulations, NIHAO has more galaxies with higher resolution over a
wider range of galaxy stellar masses $10^5 \lta \Mstar \lta 10^{11}
\Msun$.  In \S 2 we describe the observations we compare to from the
SPARC survey \citep{Lelli16}. In \S 3 we describe the NIHAO galaxy
simulations. In \S 4 we present the RAR from NIHAO and compare to
observations from SPARC. In \S 5 we investigate the origin of the
small scatter in the RAR, and present a comparison of dark matter
profiles. We finish with a summary in \S 6.

\section{SPARC Observations} \label{sec:obs}

As our observational sample we use the Spitzer Photometry and Accurate
Rotation Curves (SPARC) database \citep{Lelli16}. SPARC is the largest
sample of galaxy rotation curves with spatially resolved data on the
distribution of both stars and gas. The full sample of 175 galaxies
spans the full range of stellar masses of rotationally supported
galaxies ($10^7 \lta \Mstar \lta 10^{11}\Msun$).  It includes
near-infrared ($3.6\mu$m) observations to trace the distribution of
stellar mass and 21 cm observations that trace the atomic hydrogen gas
(\hi). The 21 cm data also provide velocity fields from which rotation
curves are derived. In some cases the rotation curves are supplemented
with higher spatial resolution rotation curves from ionized gas. All
rotation curves are nominally corrected for inclination.

Following \citet{McGaugh16} we apply some quality criteria. Ten
galaxies with inclination $i < 30$ are removed to minimize $\sin(i)$
corrections to the observed velocities. Twelve galaxies with
asymmetric rotation curves are rejected (these are flagged with
$Q=3$). This leaves a sample of 153 galaxies. Additionally, only data
points with relative velocity errors less than 10\% are kept.

\begin{figure}
\includegraphics[width=0.45\textwidth]{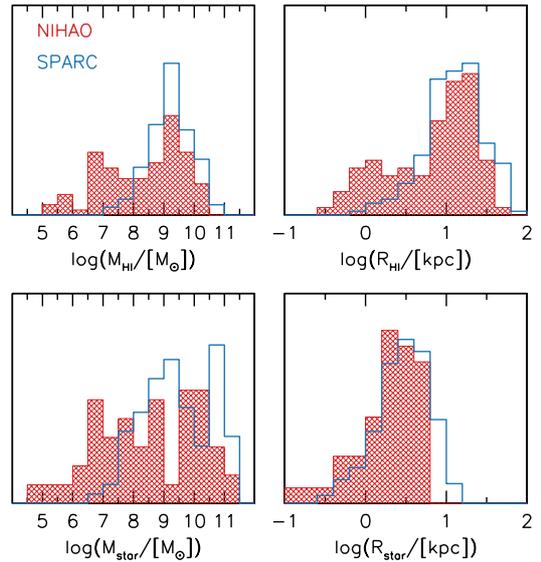}
\caption{ Histograms of masses and sizes  for atomic hydrogen
  and stars from NIHAO simulations (red, shaded), and SPARC
  observations (blue). For SPARC the stellar masses and projected
  half-light radii are obtained from $3.6\mu$m photometry, while for
  NIHAO they are calculated from the stellar particles.}
\label{fig:global}
\end{figure}

\begin{figure*}
\includegraphics[width=0.45\textwidth]{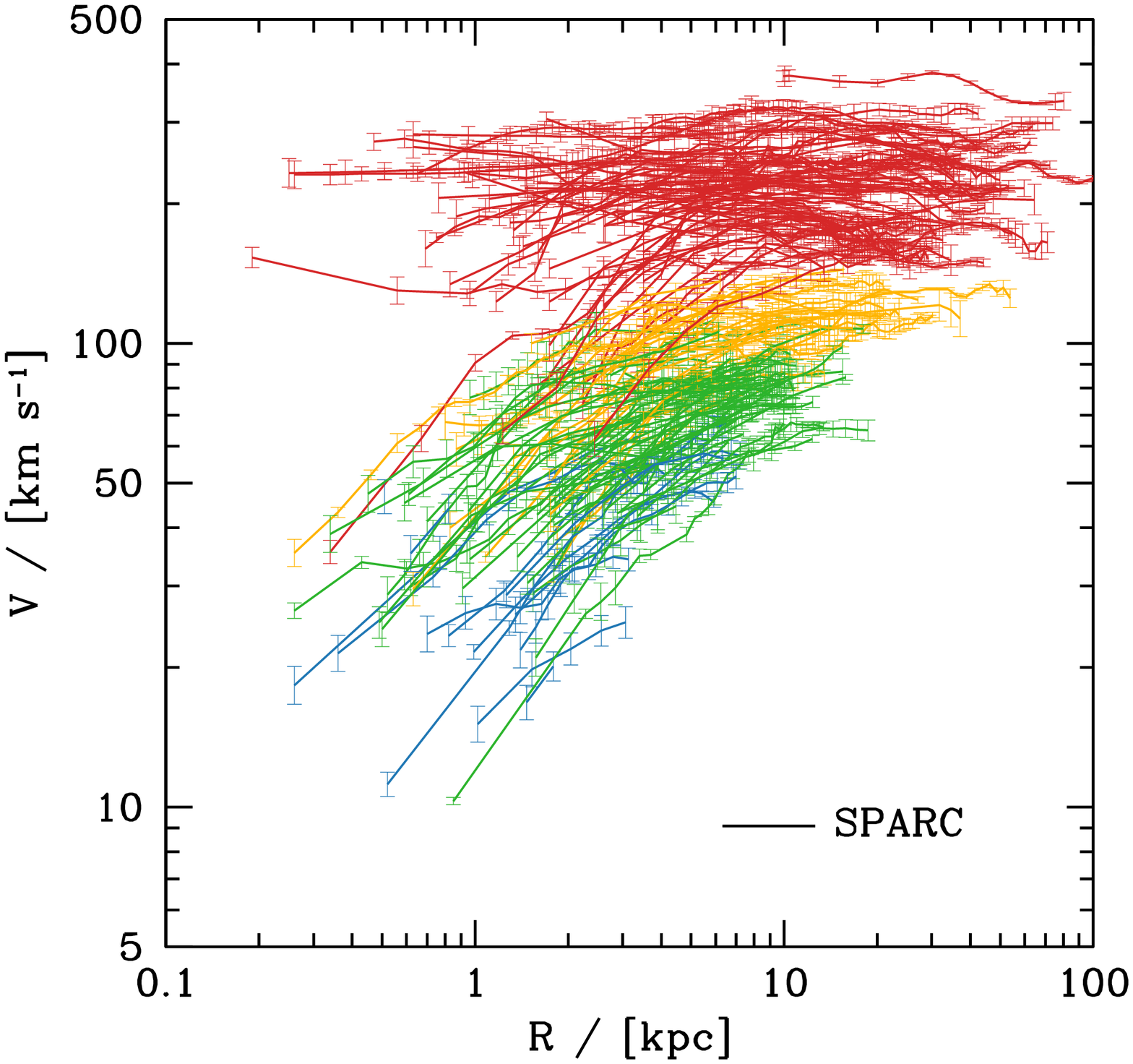}
\includegraphics[width=0.45\textwidth]{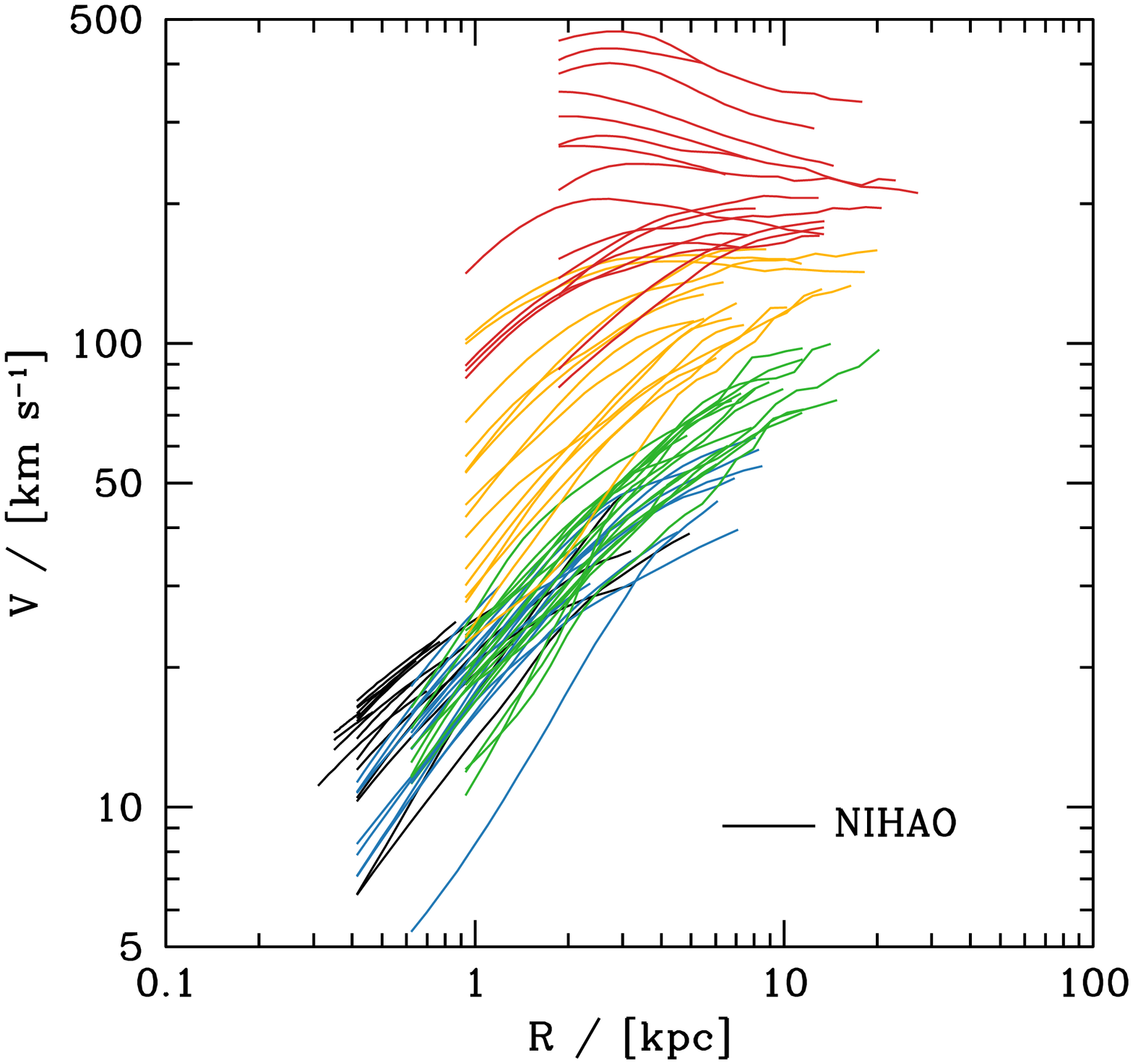}
\includegraphics[width=0.45\textwidth]{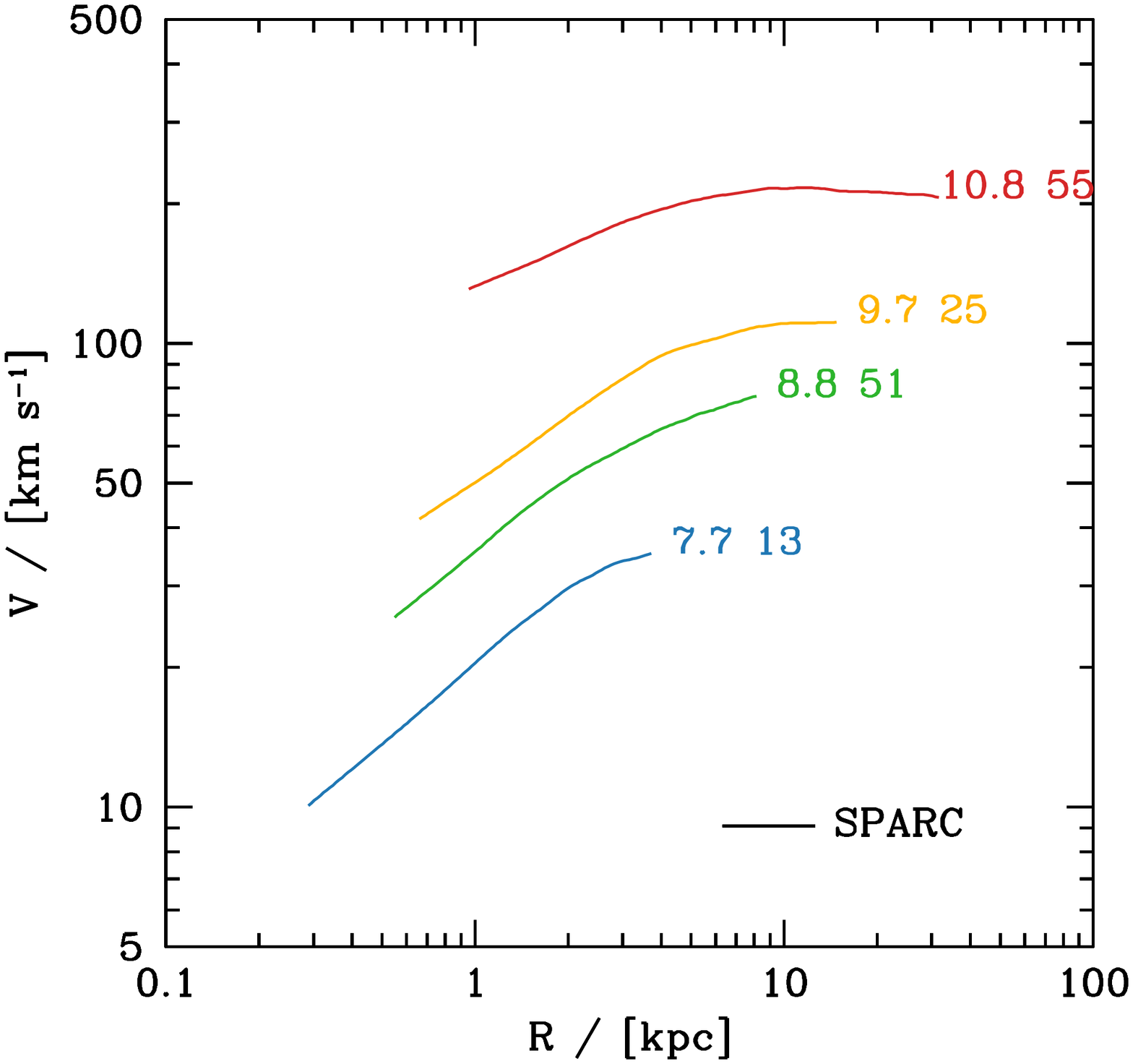}
\includegraphics[width=0.45\textwidth]{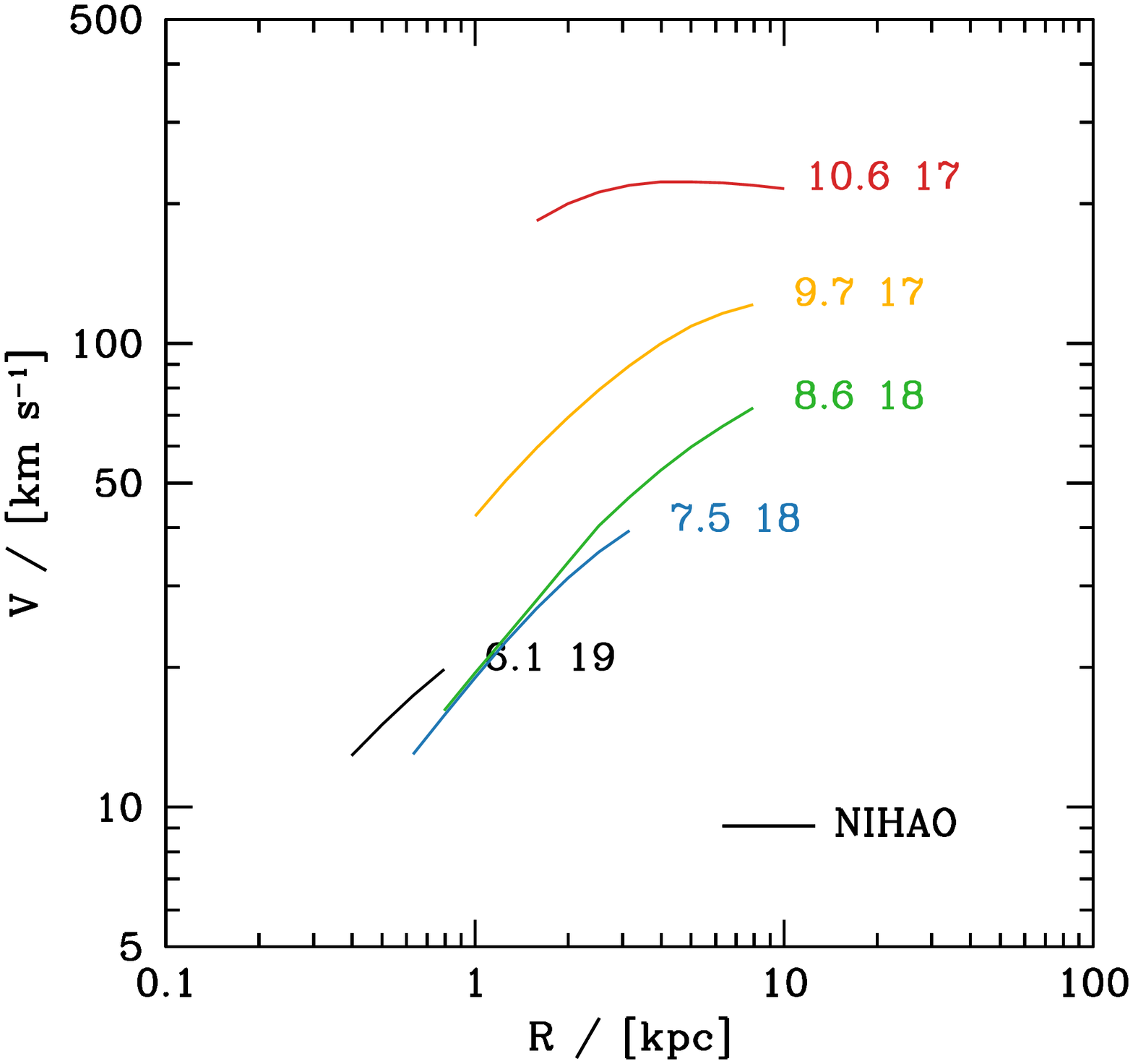}
\caption{Rotation curves from SPARC observations (left) and circular
  velocity profiles in the plane of the disk from NIHAO simulations
  (right). Upper panels show all the individual profiles color coded
  by the stellar mass. The boundaries of the mass bins are
  $\log(\Mstar/\Msun)=4.5,7.0,8.0,9.3,10.2,11.4$, and are chosen to
  have equal numbers of simulated galaxies in each bin.  For NIHAO
  these are shown from twice the force softening of the dark matter
  particles to the \hi\, radius (which encloses 90\% of the \hi\,
  mass). Lower panels show mean profiles in bins of stellar mass, and
  are plotted between the average minimum and maximum radii of the
  profiles. The mean stellar masses are shown to the right of each
  line (this can be slightly different between NIHAO and SPARC due to
  the different distribution of galaxies within each mass bin).}
\label{fig:rotcurve}
\end{figure*}

Fig.~\ref{fig:global} shows histograms of the global stellar and
atomic hydrogen gas properties of the SPARC galaxies (blue lines), and
NIHAO simulations (red shaded, see \S\ref{sec:sims}): Stellar mass,
$\Mstar$; half-light radius at 3.6 micron, $R_{\rm star}$; atomic
hydrogen mass, $M_{\rm HI}$; and radius at which \hi\, density equals
1 [$\Msun $pc$^{-2}$], $R_{\rm HI}$.  The upper left panel of
Fig.~\ref{fig:rotcurve} shows the rotation curves of this sample,
where the color code corresponds to the stellar mass. The boundaries
of the mass bins are $\log(\Mstar/\Msun)=6.7,8.0,9.3,10.2,11.4$.  The
lower panel shows average rotation curves  in bins of stellar
mass. For each mass bin we interpolate each rotation curve onto a
radial grid, then we average in $\log(V)$. We also calculate the
average minimum and maximum radius, which defines the range over which
the average rotation curves are plotted.  The mean mass is given to
the right of each average rotation curve. We see the shape and extent
of the rotation curve systematically changes from low to high mass.

Here we briefly discuss uncertainties in the observations.
\begin{enumerate}
\item Conversion from stellar light to stellar mass. The nominal value
  is assumed to be $M/L_{3.6}=0.5$ for the disk and $0.7$ for the
    bulge \citep{McGaugh16}, but could plausibly vary from 0.2 to 1.0
  depending on initial mass function (IMF), age and metallicity
  \citep{McGaugh15}. 

\item Distance. Since these galaxies are nearby, distance errors can
  be significant (up to 30\%). Only a subset of the SPARC sample has
  accurate (5\%) distances from e.g., using the Tip of the Red Giant
  Branch (TRGB).  Since physical size scales as $D$, and luminosity as
  $D^2$, the acceleration of the baryons ($\sim\,M_{\rm bar} / r^2$) is
  independent of distance.  The observed rotation velocity is also
  independent of distance (ignoring beam smearing which is worse for
  more distant galaxies), so that the total acceleration depends on
  distance as $D^{-1}$.

\item Rotation curves. The conversion from observed rotation to
  circular velocity depends on inclination, pressure support and warps
  (especially at large radii).

\item Gas budget. The majority of gas in the galaxy is in atomic
  hydrogen, which is observed for SPARC galaxies, but there is also a
  small amount of molecular and ionized gas, which is not easily
  observed, and generally missing from the SPARC data. 
\end{enumerate}

\begin{figure*}
  \includegraphics[width=0.45\textwidth]{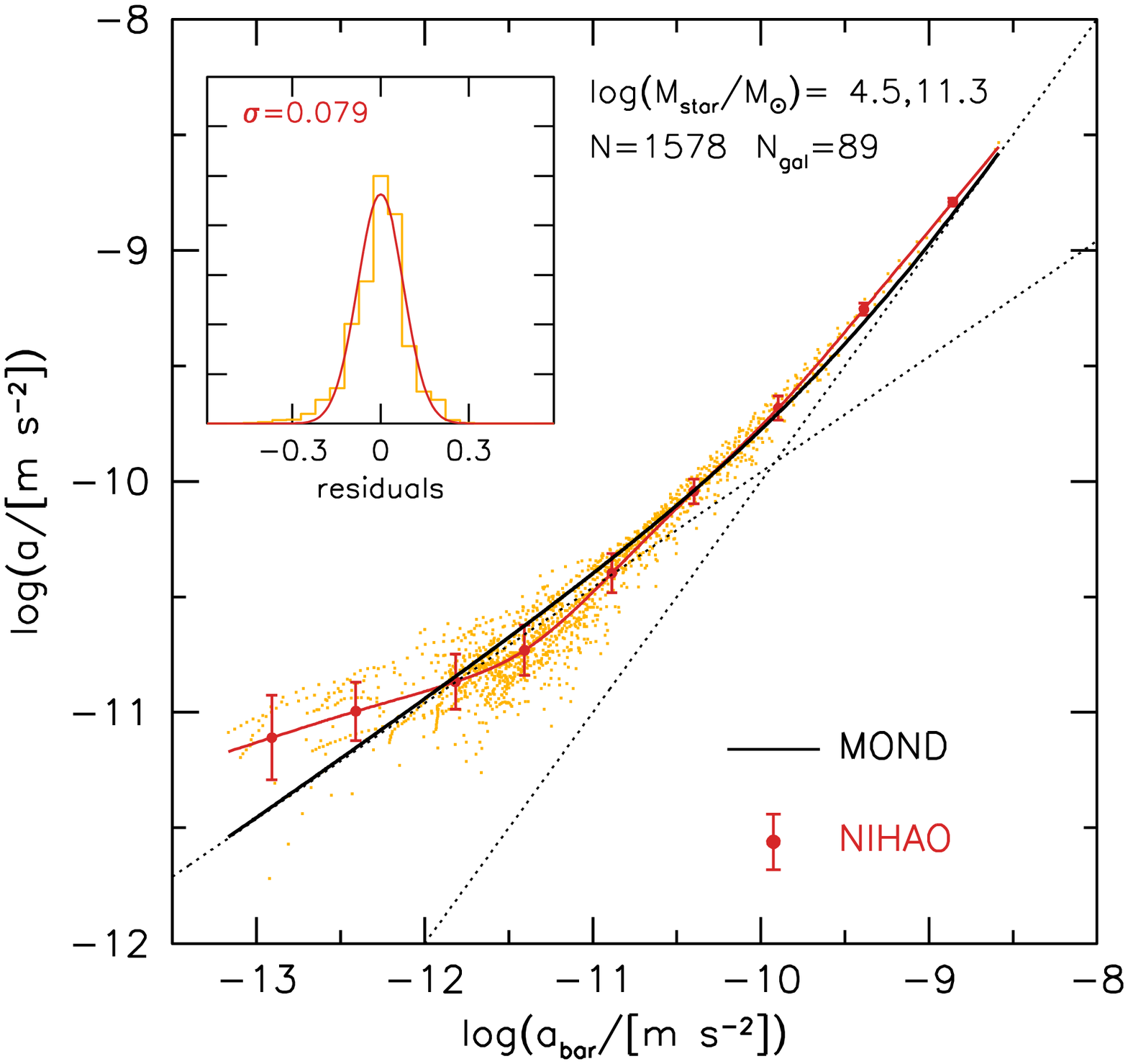}
  \includegraphics[width=0.45\textwidth]{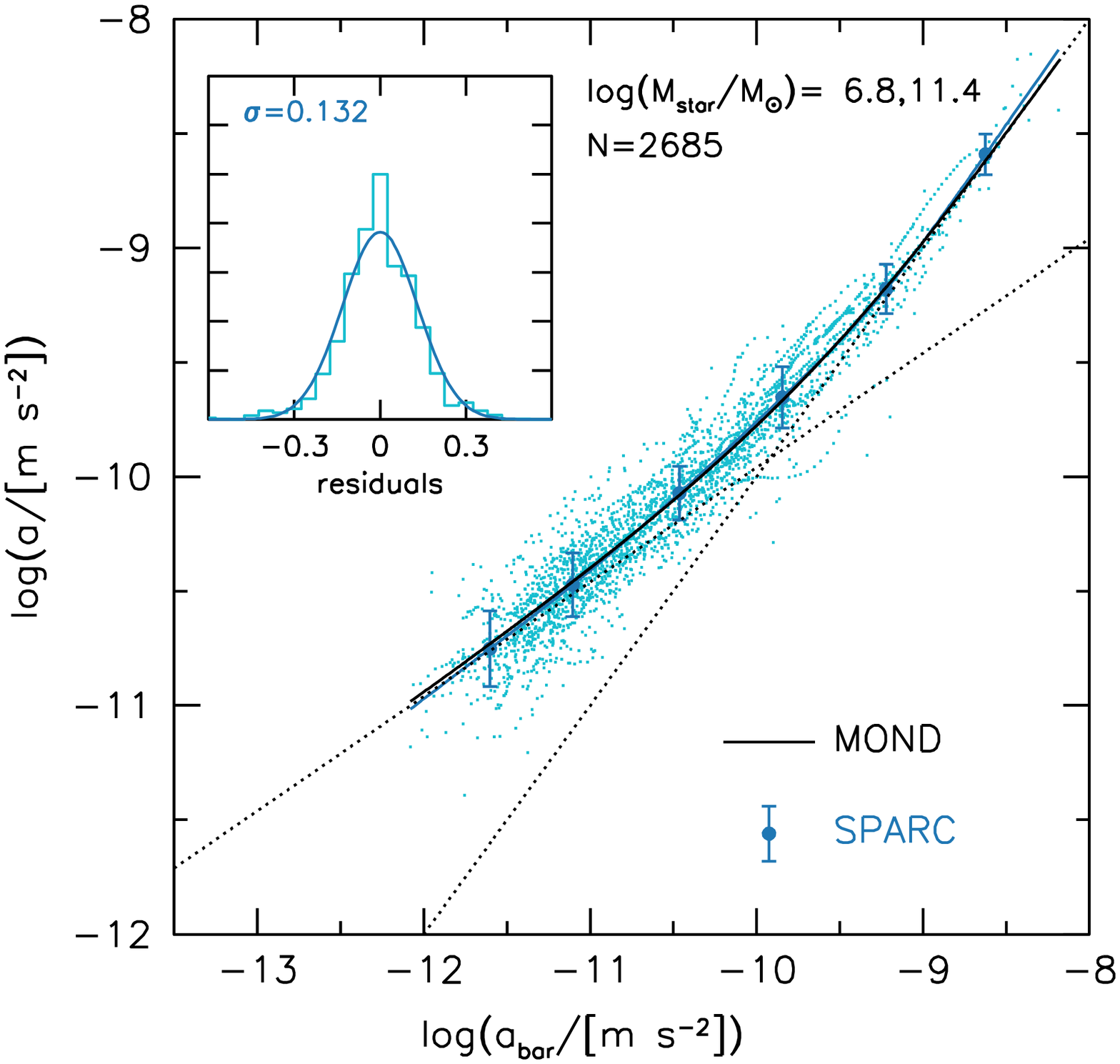}
\caption{Total acceleration, $a$, vs acceleration due to baryons,
  $a_{\rm bar}$, in NIHAO simulations (left red/orange) and SPARC
  observations (right, blue/cyan) compared to MOND (black solid line)
  with an acceleration scale $a_0=1.2\times 10^{-10}$m s$^{-2}$. 
  Small points show the individual data points. Large circles with
  error bars show the mean and $1\sigma$ scatter in bins of $a_{\rm
    bar}$. The lines show the spline interpolated mean relations.  The inset
  panel shows the residuals (orange/cyan histogram) relative to the
  interpolated mean relations, which has a standard deviation of just
  0.079 dex in NIHAO and 0.132 dex in SPARC. The corresponding
  Gaussian is shown in red (for NIHAO) and blue (for SPARC).
  The range of galaxy stellar masses, $\Mstar$, is as indicated, while
  $N$ is the number of data points.}
\label{fig:aabar}
\end{figure*}

\section{NIHAO Galaxy Formation Simulations} \label{sec:sims}

For theoretical predictions for the RAR in a \LCDM universe we use the
NIHAO galaxy formation simulations \citep{Wang15}. These are a sample
of $\sim 90$ cosmological hydrodynamical galaxy formation simulations
run with the SPH code {\sc gasoline2} \citep{Wadsley17}. 

Haloes are selected at redshift $z=0$ from parent dissipationless
simulations of size 60, 20, \& 15 Mpc$/h$, presented in
\citet{Dutton14} which adopt a flat $\Lambda$CDM cosmology with
parameters from the \citet{Planck14}: Hubble parameter $H_0$= 67.1
\kms Mpc$^{-1}$, matter density $\Omegam=0.3175$, dark energy density
$\Omega_{\Lambda}=1-\Omegam=0.6825$, baryon density $\Omegab=0.0490$,
power spectrum normalization $\sigma_8 = 0.8344$, power spectrum slope
$n=0.9624$.  Haloes are selected uniformly in log halo mass from $\sim
10$ to $\sim 12$ {\it without} reference to the halo merger history,
concentration or spin parameter.  Star formation and feedback is
implemented as described in \citet{Stinson06, Stinson13}.  Mass and
force softening are chosen to resolve the mass profile of the
  target halo at $\lta 1\%$ the virial radius, which results in $\sim
10^6$ dark matter particles inside the virial radius of all haloes at
$z=0$. The motivation of this choice is to ensure that the simulations
resolve the galaxy dynamics on the scale of the half-light radii,
which are typically $\sim 1.5\%$ of the virial radius
\citep{Kravtsov13}.

Each hydro simulation has a corresponding dark matter only (DMO)
simulation of the same resolution. These simulations have been started
using the identical initial conditions, replacing baryonic particles
with dark matter particles. The full sample of hydro simulations
  we use is 89. When comparing hydro to DMO we remove 5 simulations
for which either the hydro or DMO simulation is undergoing a major
merger and is thus out of equilibrium.

Haloes in NIHAO zoom-in simulations were identified using the
MPI+OpenMP hybrid halo finder
\texttt{AHF}\footnote{http://popia.ft.uam.es/AMIGA} \citep{Gill04,
  Knollmann09}. \texttt{AHF} locates local over-densities in an
adaptively smoothed density field as prospective halo centers.  In
this study we use one halo from each zoom-in simulation (the one with
the most particles).  The virial masses of the haloes are defined as
the masses within a sphere whose average density is 200 times the
cosmic critical matter density, $\rhocrit=3H_0^2/8\pi G$.  The virial
mass, size and circular velocity of the hydro simulations are denoted:
$M_{200}$, $R_{200}$, $V_{200}$.  The corresponding properties for the
dark matter only simulations are denoted with a superscript, ${\rm
  DMO}$.  For the baryons we calculate masses enclosed within spheres
of radius $r_{\rm gal}=0.2R_{200}$, which corresponds to $\sim 10$ to
$\sim 50$ kpc.  The stellar mass inside $r_{\rm gal}$ is $M_{\rm
  star}$, the atomic hydrogen, \hi, inside $r_{\rm gal}$ is computed
following \citet{Rahmati13} as described in \citet{Gutcke17}.

The NIHAO simulations are the largest set of cosmological zoom-ins
covering the halo mass range $10^{10} \lta M_{200} \lta 10^{12}\Msun$. Their
uniqueness is in the combination of high spatial resolution coupled to
a statistical sample of haloes. In the context of \LCDM they form the
``right'' amount of stars both today and at earlier times
\citep{Wang15}. Their cold gas masses and sizes are consistent with
observations \citep{Stinson15, Maccio16, Dutton19}, they follow the
gas, stellar, and baryonic Tully-Fisher relations \citep{Dutton17}. On
the scale of dwarf galaxies the dark matter haloes expand yielding
cored dark matter density profiles consistent with observations
\citep{Tollet16}, and resolve the too-big-to-fail problem of field
galaxies \citep{Dutton16}.  They reproduce the diversity of dwarf
galaxy rotation curve shapes \citet{Santos-Santos18}, and the \hi\,
linewidth velocity function \citep{Maccio16, Dutton19}. They match the
observed clumpy morphology of galaxies seen in CANDELS \citep{Buck17}.
As such they provide a good template with which to study the RAR in a
\LCDM context. 

Histograms of the stellar and atomic hydrogen masses and sizes are
shown with red shading in Fig.~\ref{fig:global}. Broadly speaking
NIHAO and SPARC have similar distributions, but NIHAO extends to lower
masses. The right panels of Fig.~\ref{fig:rotcurve} shows the circular
velocity profiles from the NIHAO galaxy formation simulations. These
have been calculated using the potential in the plane of the disk
using Eq.1.  The NIHAO simulations have been plotted from two dark
matter force softening lengths ($\approx$ the convergence radius), to
the radius enclosing 90\% of the HI ($\approx$ the edge of the
observable galactic HI).  The radial range of the velocity profiles is
similar, but slightly narrower.  The change in velocity profile shapes
from low to high mass is similar between simulations and
observations. In detail there are differences, caused at least in part
by differences in the distribution of baryons.

\begin{figure}
\centering{\includegraphics[width=0.45\textwidth]{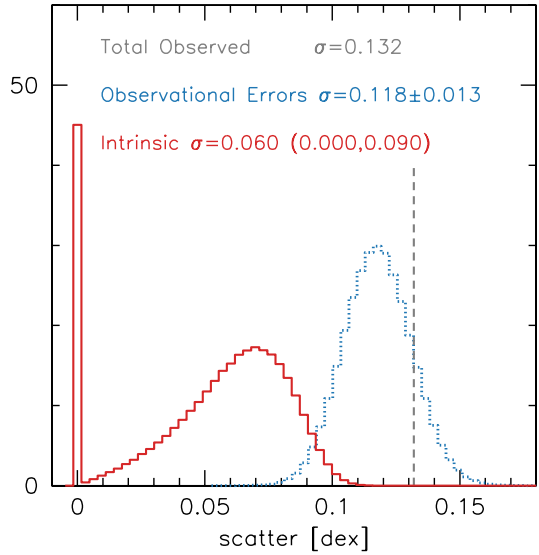}
  }
\caption{Scatter in the RAR. The total observed scatter is
  $0.132$ (grey dashed vertical line). The observational errors
  contribute $0.118\pm0.013$ (blue dotted
  histogram). Subtracting the observational errors gives the intrinsic
  scatter (red histogram). For cases where the observed error is
  greater than the total scatter we set $\sigma= 0$, which is the
  spike. The intrinsic scatter has a median of 0.06 and a 90\%
  confidence interval of 0.00 to 0.09. }
\label{fig:rar_scatter}
\end{figure}

\section{The radial acceleration relation} \label{sec:mdar}

Fig.~\ref{fig:aabar} shows the relation between total acceleration,
$a$, and the acceleration due to the baryons, $a_{\rm bar}$ for NIHAO
galaxies (left) and SPARC observations (right). The dots show all of
the individual measurements for observed and simulated galaxies. The
total number of data points, $N\sim 1400$ for NIHAO and $N \sim 2700$
for SPARC. For NIHAO we sample the circular velocity profiles linearly
in units of the virial radius, but we only include points which are
observable (i.e., within the HI radius). This is similar to the
observations, which tend to sample the rotation curves linearly in
radius, and with similar numbers of data points for different mass
galaxies. The large points with error bars show the mean of
$\log_{10}(a)$ in bins of $\log_{10}(a_{\rm bar})$, while the error
bar shows the scatter about the spline interpolated mean relation. 

The solid black line shows the relation fitted to SPARC data by
\citet{McGaugh16}. The functional form was chosen to follow the
asymptotic relations for high and low accelerations in MOND (dotted
lines). Overall the NIHAO simulations provide a very good match to the
observed RAR. The agreement is especially good when one considers that
there are observational systematics, such as the stellar mass-to-light
ratio and distance scale that can shift the observed relation
\citep{McGaugh16, Lelli17}. Furthermore, the NIHAO simulations are
necessarily an approximation for how galaxies form in a \LCDM
universe.

\begin{figure}
  \includegraphics[width=0.45\textwidth]{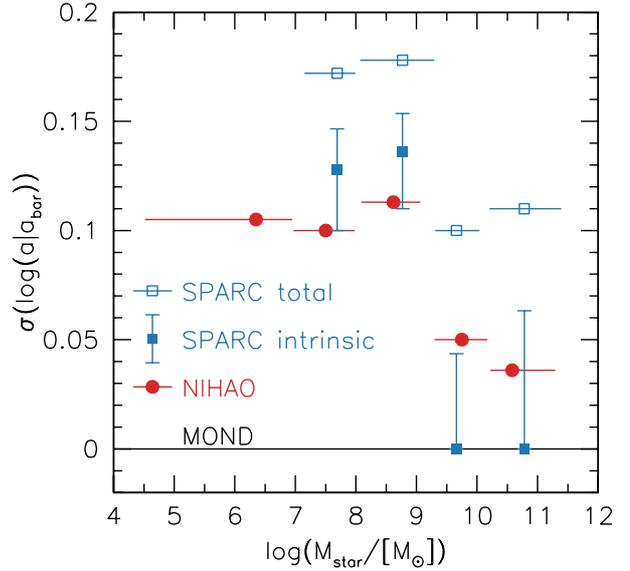}
\caption{Mass dependence of scatter in the RAR. Red circles show
  results from NIHAO simulations. Blue squares show the total and
  intrinsic scatter from SPARC galaxies. The vertical error bars show
  the intrinsic scatter from SPARC adopting measurement errors of 0.09
  and 0.14 dex, for the upper and lower limit, respectively. The
  horizontal lines show the range of $\Mstar$ in each bin.}  
\label{fig:rar_scatter_mass}
\end{figure}

The inset panels show the scatter about the interpolated mean relation
(histogram). The standard deviation and corresponding Gaussian
is given in red (for NIHAO) and blue (for SPARC).
The scatter from NIHAO is just 0.079 dex. Similar
small scatters were reported by other \LCDM simulation studies
\citep{Keller17, Ludlow17}.  This is less than the total observed
scatter of 0.132 dex. However, there are observational errors which
means the intrinsic scatter is lower. Note one should be aware that in
principle observational errors can correlate with offsets from the RAR
leading to lower observed than intrinsic scatter. For example if
galaxies with higher stellar $M/L$ have lower $a$, this would result
in the RAR measured with constant $M/L$ having less scatter than the
true RAR.

\begin{figure*}
  \includegraphics[width=0.45\textwidth]{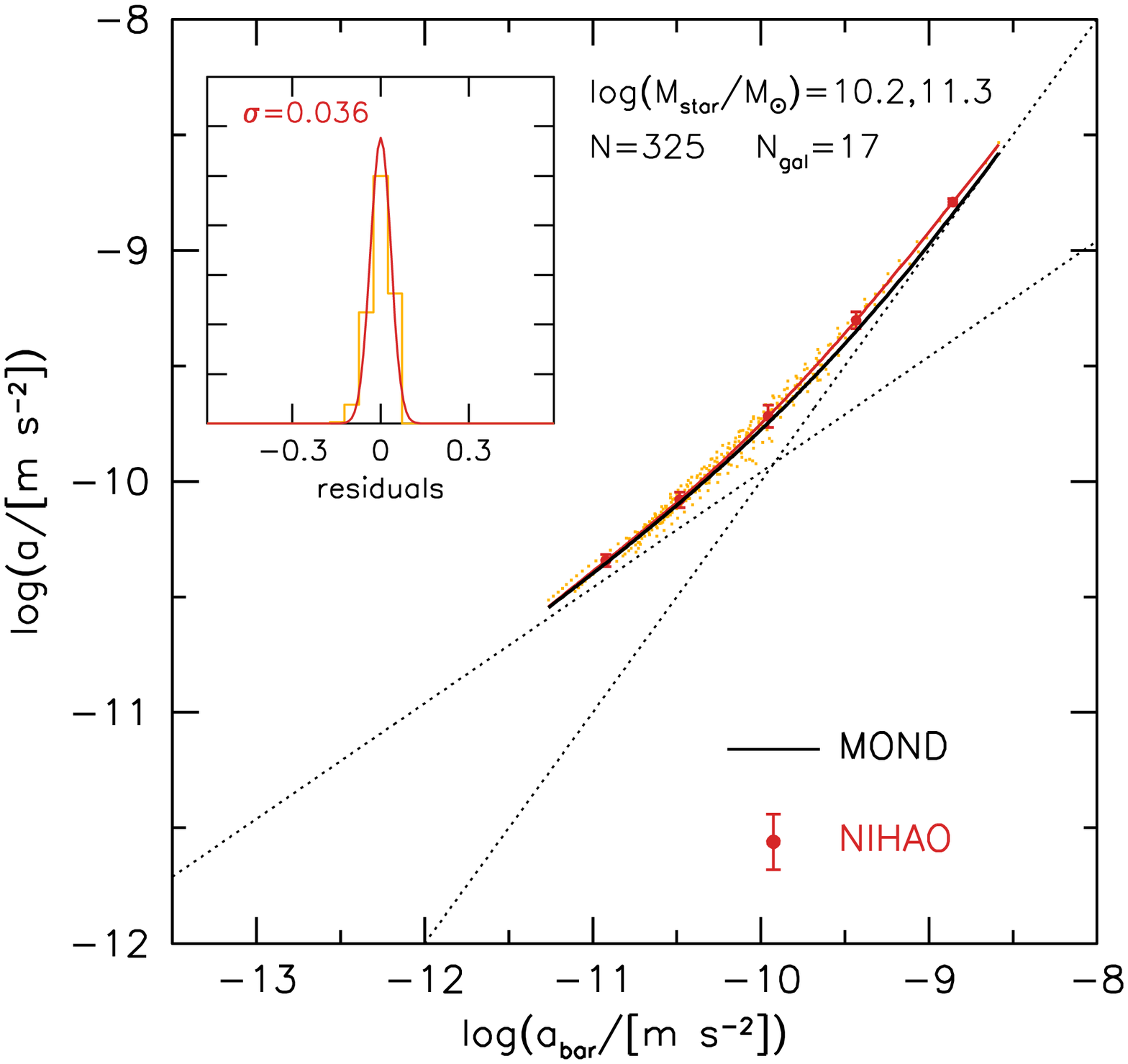}
  \includegraphics[width=0.45\textwidth]{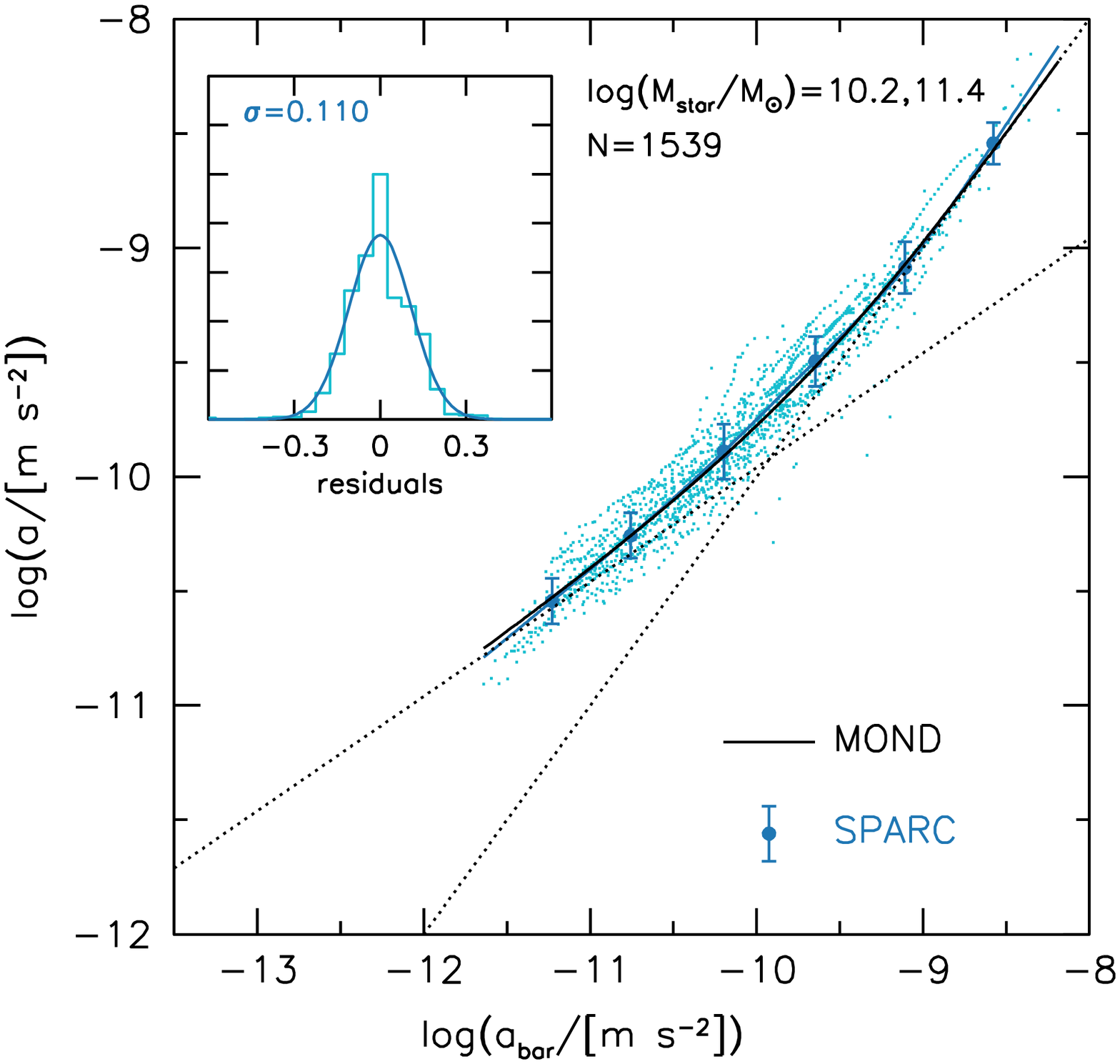}
  \includegraphics[width=0.45\textwidth]{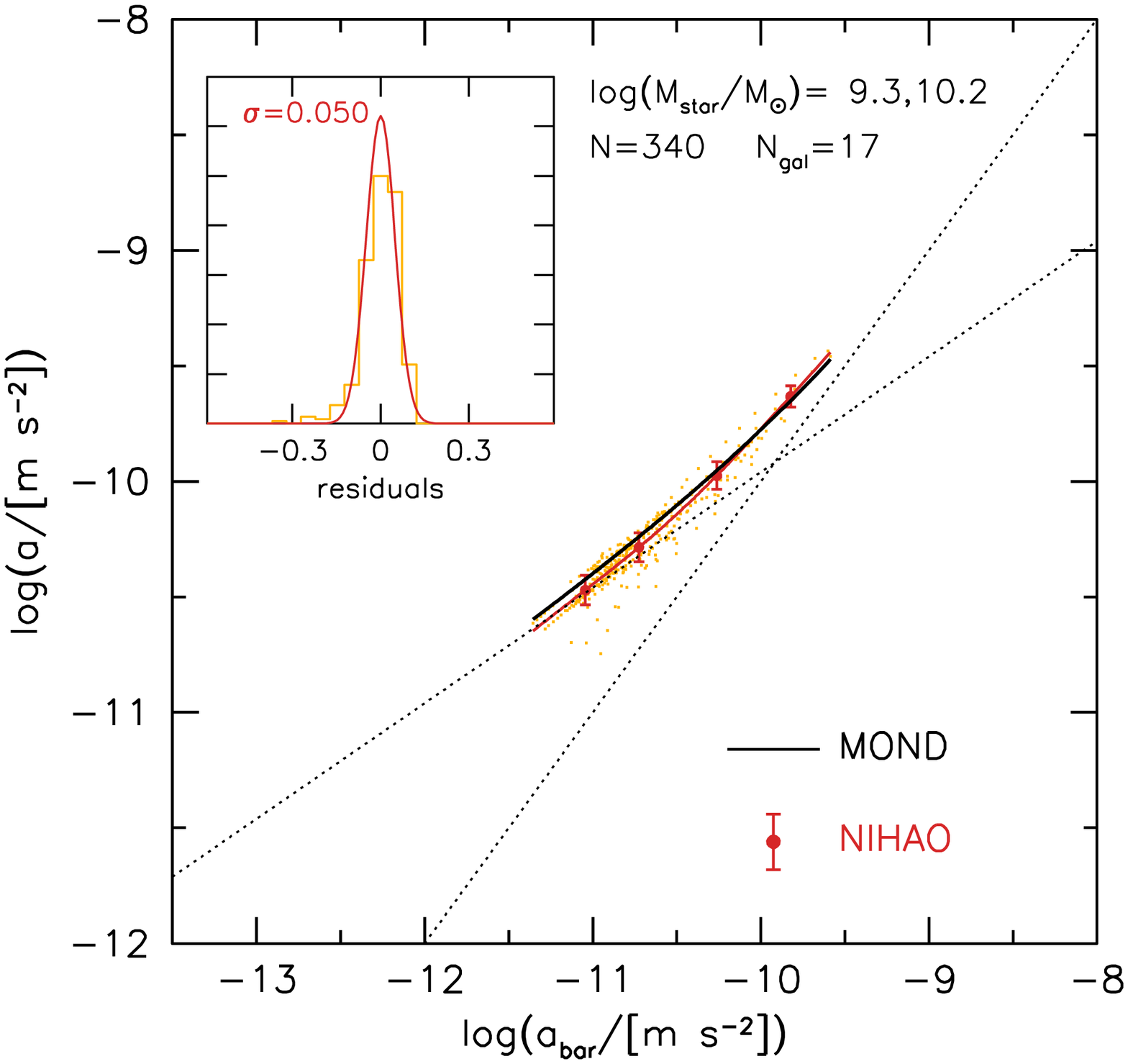}
  \includegraphics[width=0.45\textwidth]{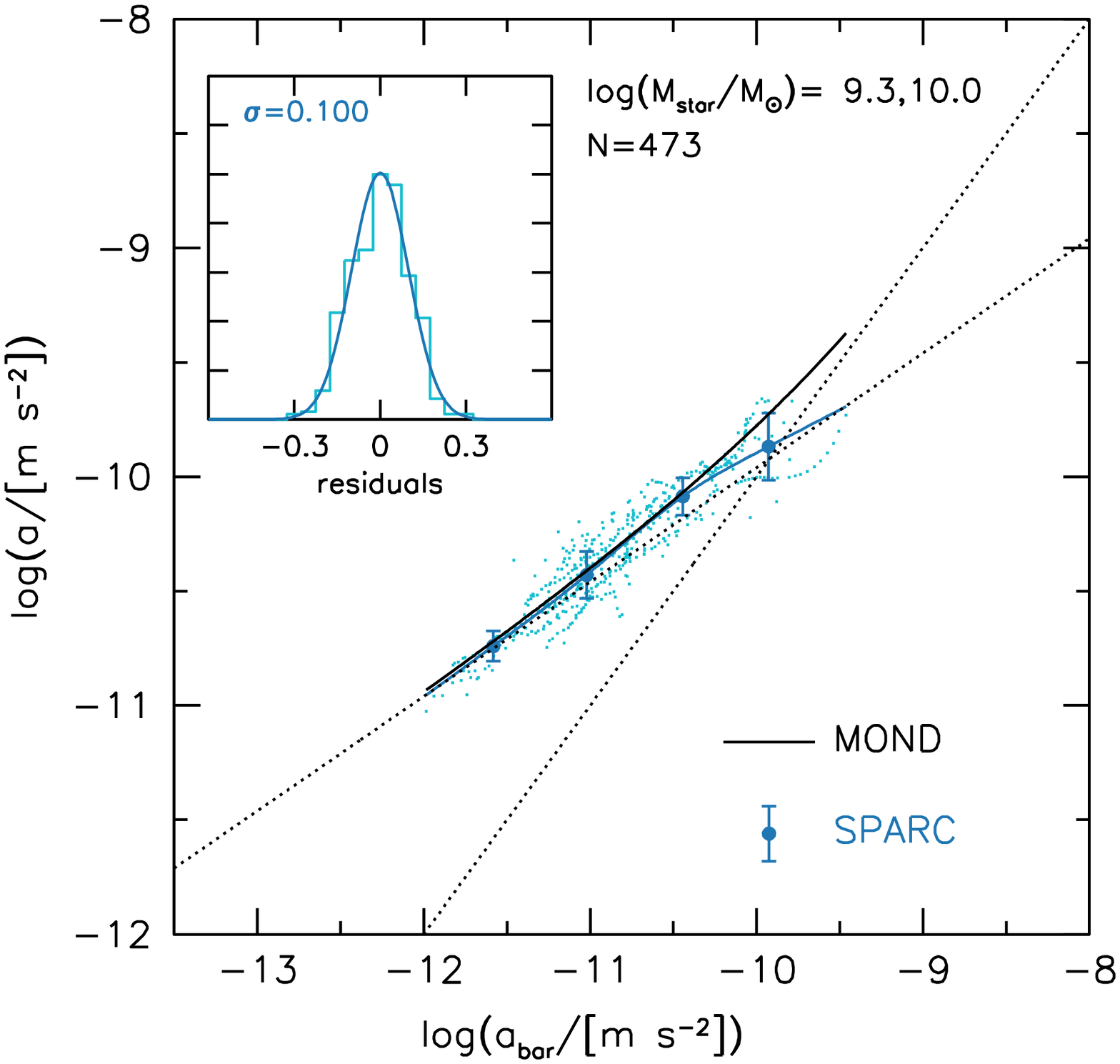}
\caption{As Fig.~\ref{fig:aabar} but for galaxies in narrow stellar
  mass ranges as indicated at the top right of each panel. At these
  (high) masses the simulated RAR is very similar to the observed one,
  and has very small scatter making the individual data points are
  hard to see.}
\label{fig:aabar2}
\end{figure*}

\begin{figure*}
  \includegraphics[width=0.45\textwidth]{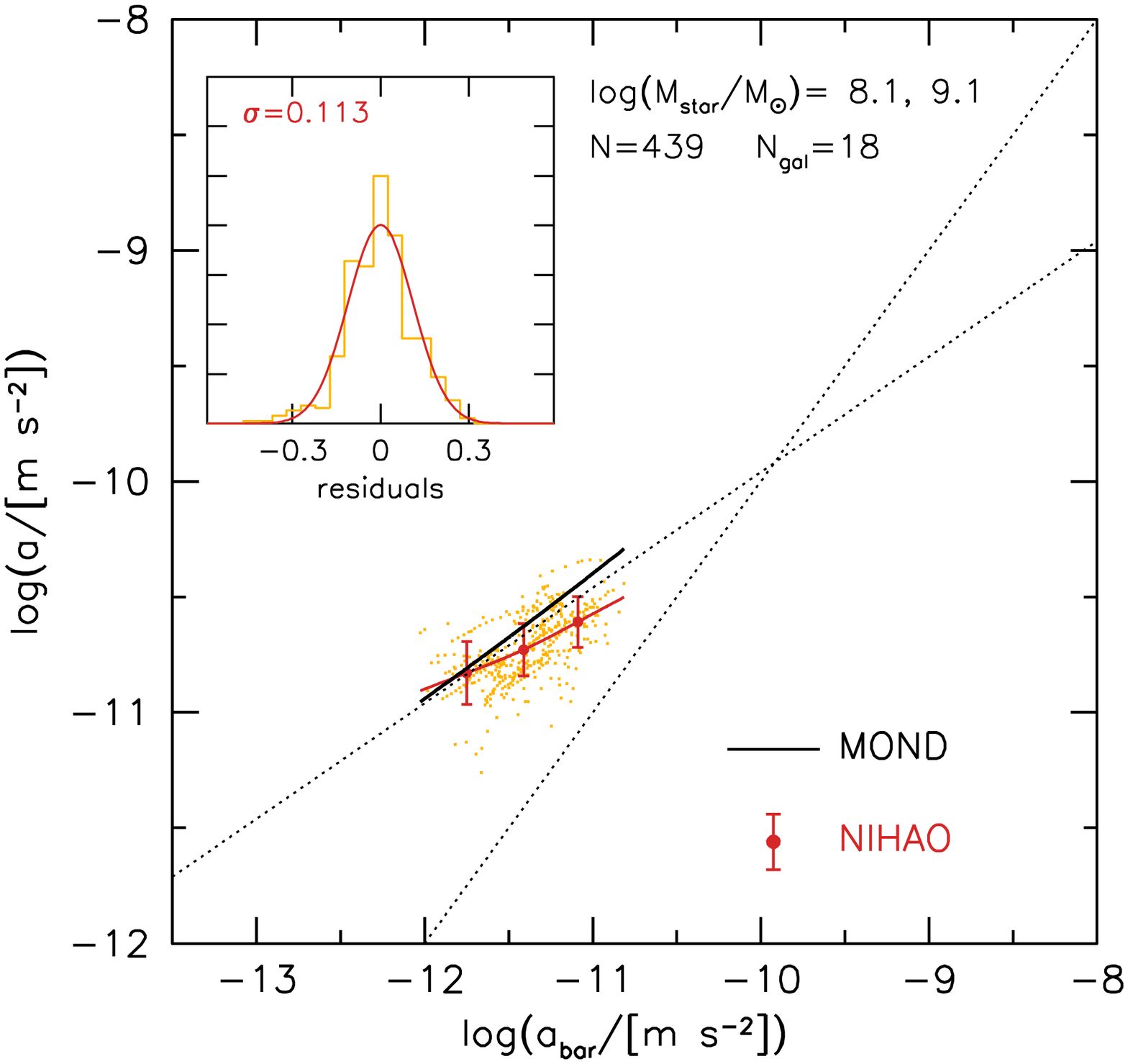}
  \includegraphics[width=0.45\textwidth]{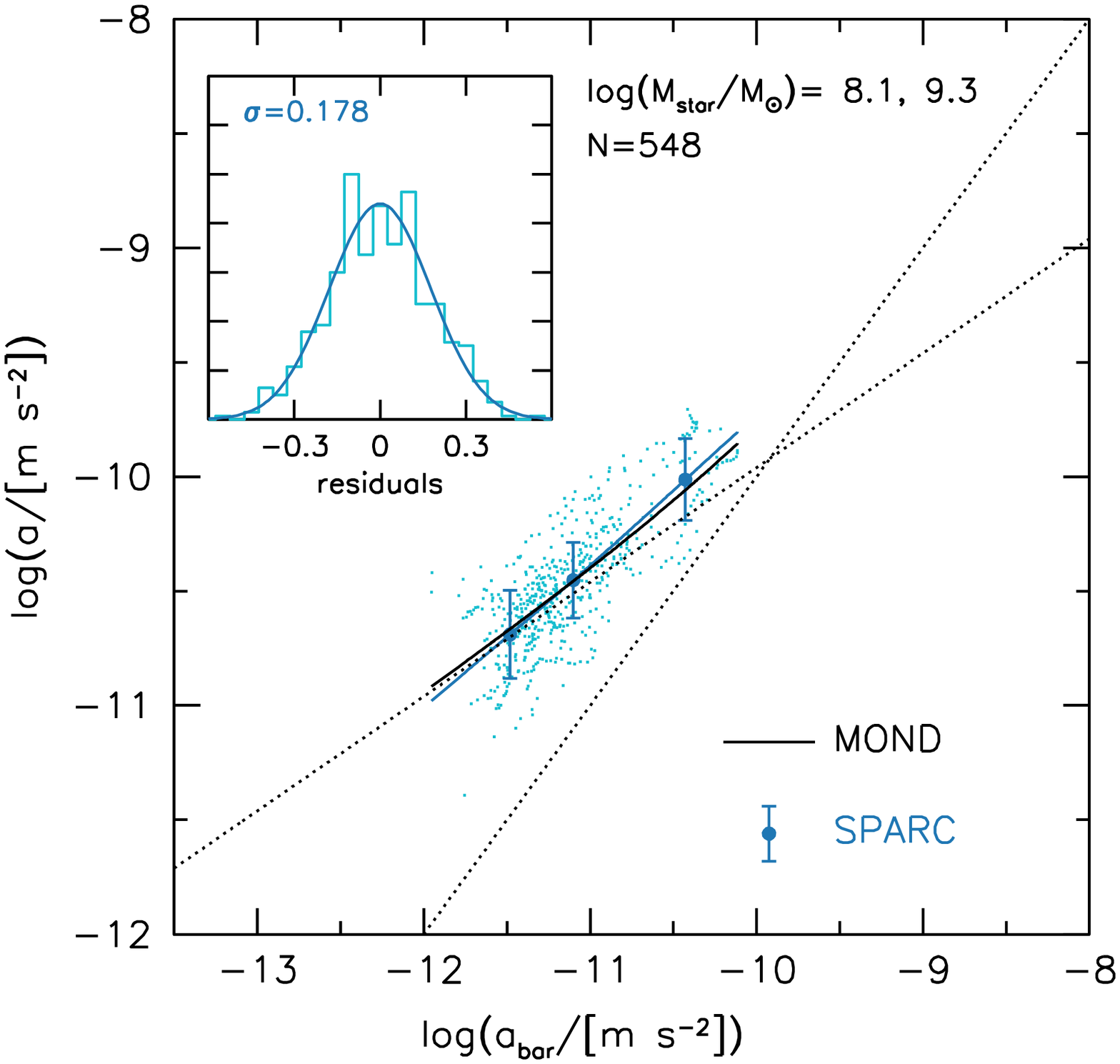}
  \includegraphics[width=0.45\textwidth]{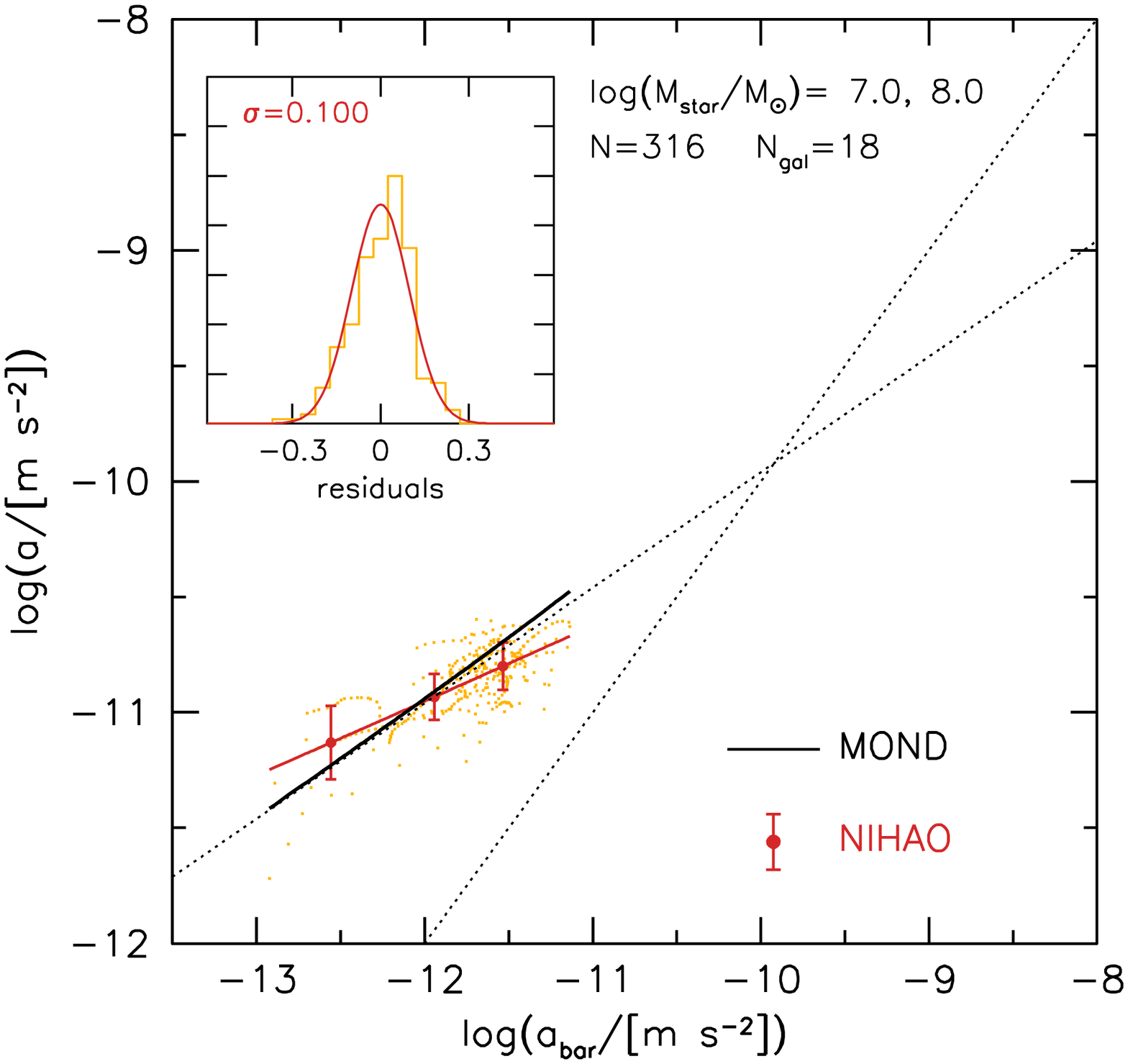}
  \includegraphics[width=0.45\textwidth]{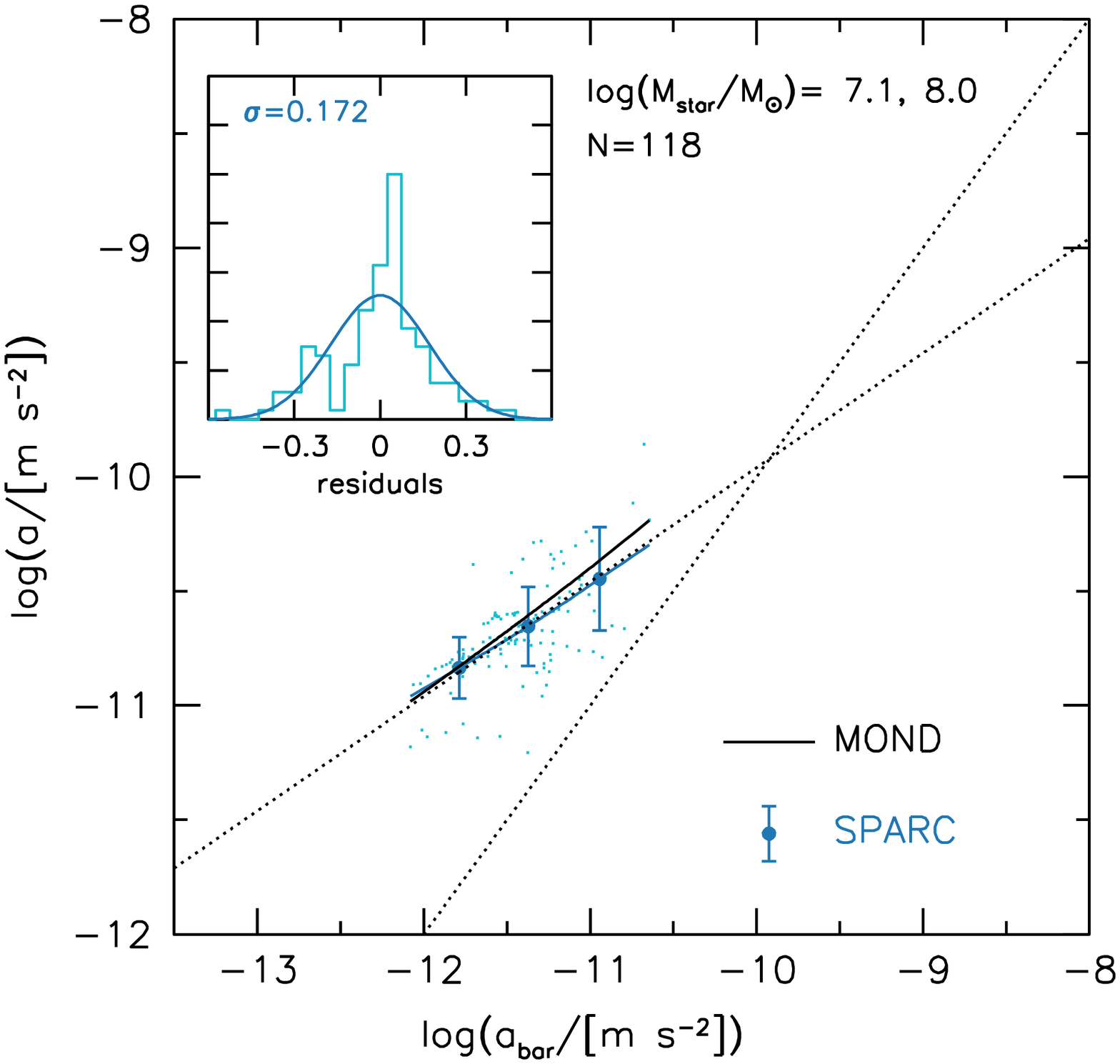}
  \caption{As Fig.~\ref{fig:aabar} but for galaxies in narrow stellar
  mass ranges as indicated at the top right of each panel. At these
  (low) masses the simulated RAR shows small but significant
  deviations from the observed one. Both simulations and observations
  have larger scatter than at higher masses.}
\label{fig:aabar3}
\end{figure*}

\subsection{Intrinsic scatter from observations}
As discussed above, observational sources of error include distance,
stellar mass-to-light ratio, disk inclination, rotation velocity, and
atomic hydrogen flux. According to \citet{McGaugh16} these contribute
0.08, 0.06, 0.05, 0.03, 0.01 dex to the scatter in $a$ at $a_{\rm
  bar}$, yielding a total observational error of 0.116
dex. Subtracting this (in quadrature) from the total observed scatter
of 0.132 dex yields an intrinsic scatter of 0.063 dex. 

However, it should be noted that the observational errors are just
estimates. The true observational errors could be larger or
smaller. For example, \citet{McGaugh16} adopt an error on the stellar
mass-to-light ratio of 0.11 dex. The true error could plausibly be
anywhere in the range 0.05 to 0.20 dex, and it could depend
systematically on the galaxy mass, or other galaxy property.  To get
an idea of the plausible range of the intrinsic scatter we assume the
observational errors are drawn from a Gaussian with the mean specified
above, and a standard deviation of 20\% of the mean.  For example, if
the distance errors are reported to be are 0.08 dex, we adopt a
standard deviation of 0.016 dex. 

Assuming the uncertainties in the errors of the five sources are
uncorrelated, using a Monte Carlo simulation the total observational
error is $0.118 \pm 0.013$ (Fig.~\ref{fig:rar_scatter}). Subtracting
this from the total observed scatter of 0.132 dex yields an intrinsic
scatter with a wide variation (red histogram). The 90\% confidence
interval ranges from zero to $0.090$ dex. Thus the scatter in the RAR
from our simulations of 0.079 dex is consistent with the
observations. The prediction from MOND for zero intrinsic scatter is
also consistent with the observations. In order to use the scatter in
the RAR to distinguish between \LCDM and MOND, either a more accurate
measurement is required, or as we shall see below we need to look at
the RAR scatter for different mass galaxies.

The largest source of observational uncertainty are the distances,
which account for half of the total variance. So if these can be
reduced a significant improvement in the intrinsic scatter is
possible. The galaxies with the largest distance errors from the SPARC
survey use the Hubble flow, $D=v_{\rm sys}/H_0$, so the unknown
peculiar velocities provide a random source of error, while the value
of the Hubble constant provides a systematic error. The peculiar
velocity error can be reduced by simply observing galaxies at larger
distances.  This is easier said than done, because galaxies that are
further away have smaller angular sizes. It is possible to obtain
rotation curves using optical emission lines for such galaxies
\citep[e.g.,][]{Courteau07}, but atomic hydrogen gas maps will
require future radio telescopes such as the Square Kilometer Array
(SKA). These will have the resolution and sensitivity to measure HI
density maps and rotation curves of galaxies at large enough distances
such that peculiar velocity errors are negligible.

\subsection{Mass dependence of the RAR}
The dependence of the scatter in the RAR on stellar mass is shown in
Fig.~\ref{fig:rar_scatter_mass}. In both simulations and observations
higher mass galaxies have smaller scatter than lower mass galaxies. In
the highest mass bin centered on $\Mstar\simeq 10^{10.7}\Msun$, the
scatter in NIHAO is just 0.036 dex (0.045 dex relative to MOND),
compared with 0.110 dex for SPARC. In the mass bin centered on
$\Mstar\simeq 10^{7.5}\Msun$ the scatter in NIHAO increases to 0.10
dex (0.14 dex relative to MOND), compared with 0.172 dex for SPARC.

The quadratic differences between the simulated and observed scatter
from high to low mass are 0.104, 0.087, 0.138, and 0.140 dex. In other
words, these are the size of the measurement errors needed to
reconcile the observed scatter with the simulated scatter. Based on
Fig.~\ref{fig:rar_scatter} and the reported measurement errors for
SPARC galaxies, which do not vary significantly with galaxy mass,
these are reasonable numbers. So we conclude that the mass dependent
scatter in the NIHAO RAR is consistent with the intrinsic scatter from
observations.  The zero scatter assumed by MOND is also consistent
with the high mass galaxies $10^{9.3} \lta \Mstar \lta
10^{11.4}\Msun$, but inconsistent with the low mass galaxies $10^{7.0}
\lta \Mstar \lta 10^{9.3}\Msun$. Thus using the same data that has
been used as evidence in favor of MOND by \citet{McGaugh16}, we have
shown that the intrinsic scatter in the RAR  actually disfavors MOND,
and is in excellent agreement with predictions from $\LCDM$.  Our
result echos that of \citet{Rodrigues18} who concluded that the
observed rotation curves from SPARC could not be fitted with MOND
assuming a universal value of the acceleration scale, $a_0$.

Figs.~\ref{fig:aabar2} \& \ref{fig:aabar3} shows the RAR divided into
four bins of stellar mass centered on $\Mstar\simeq 10^{7.5}$,
$10^{8.8}$, $10^{10.0}$, and $10^{10.9}\Msun$. These bins are the same
as used in Fig.~\ref{fig:rotcurve}, and have been chosen to have
roughly the same number of NIHAO galaxies per bin. 

A feature that stands out in both simulations and observations is that
higher mass galaxies span a wider range of accelerations than lower
mass galaxies.  This is due to the systematic change in the shape of
the circular velocity profiles with galaxy mass (see
Fig.~\ref{fig:rotcurve}). The highest mass galaxies have velocity
profiles close to constant, and thus acceleration varies inversely
with radius, whereas low mass galaxies have rising velocity profiles
with slopes close to 0.5, which results in acceleration independent of
radius. 

For the two highest mass bins $10^{9.3}\lta \Mstar \lta
10^{11.3}\Msun$ (Fig.~\ref{fig:aabar2}) the mean NIHAO RAR follows the
MOND RAR extremely closely. For the two lowest mass bins $10^{7.0}
\lta \Mstar \lta 10^{9.3}\Msun$ (Fig.~\ref{fig:aabar3}) the mean NIHAO
RAR shows departures from the MOND RAR. At baryon accelerations of
$a_{\rm bar}\sim 10^{-11}\,{\rm m s}^{-2}$ MOND over-predicts NIHAO,
while at $a_{\rm bar}\lta 10^{-12}\,{\rm m s}^{-2}$ MOND
under-predicts NIHAO. This trend continues when we look at even lower
mass galaxies.

\begin{figure}
  \includegraphics[width=0.45\textwidth]{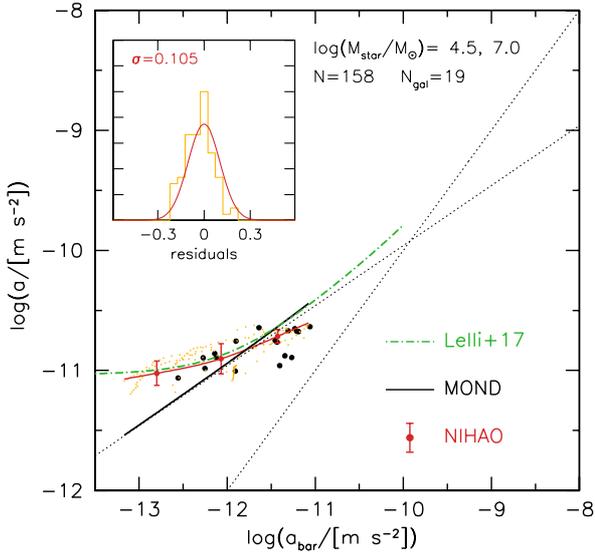}
\caption{Total acceleration, $a$, vs acceleration due to baryons,
  $a_{\rm bar}$, for the lowest stellar mass galaxies in NIHAO.  Black
  circles show the acceleration at the projected stellar half-mass
  radii. These galaxies deviate significantly from the MOND RAR, but
  are consistent with the observed relation for dwarf spheroidal
  galaxies \citep[green line, ][]{Lelli17}.}
\label{fig:aabar4}
\end{figure}

\begin{figure}
\includegraphics[width=0.45\textwidth]{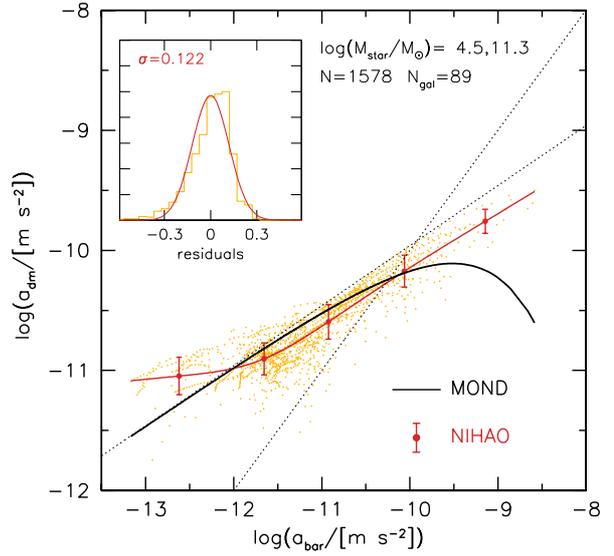}
\caption{ Dark matter acceleration, $\adm$, vs acceleration due to
  baryons, $\abar$, for NIHAO. The points with lowest and highest
  accelerations deviate the most from the MOND RAR (black line). The
  dotted lines show the 1:1 relation and the asymptotic MOND relation
  at low $\abar$.}
\label{fig:adm_abar}
\end{figure}

Fig.~\ref{fig:aabar4} shows the RAR for the lowest mass galaxies in
NIHAO, with stellar masses between $10^{4.5}\Msun$ and
$10^{7.0}\Msun$. These galaxies deviate significantly from the MOND
relation below an acceleration scale of $a_{\rm bar}=10^{-12}$ m
s$^{-2}$. The green line shows a fit to the observed RAR for dwarf
spheroidal galaxies from \citet{Lelli17}. While the observed and
simulated dwarfs have similar ranges of stellar masses, there are some
differences that should be considered. The observations are primarily
satellite galaxies, whereas the simulations are all centrals.  The
observed total accelerations are based on stellar kinematics, whereas
for the simulations we show points sampled from the radii observable
with atomic hydrogen (using the same procedure as for more massive
galaxies). The black circles in Fig.~\ref{fig:aabar4} show the RAR for
NIHAO galaxies at the location of the projected half-mass radii of the
stars.  These points tend to sample higher accelerations, but follow
the same trend as the full velocity curve points.  Even though there
are some caveats in the comparison between our \LCDM based simulations
and observations, the agreement is an encouraging sign.

The deviation of the observed RAR for dwarf Spheroidal galaxies from
the MOND prediction is another apparent failure of MOND.  The
particular assumption that breaks is that the RAR has a slope of 1/2
at low baryon accelerations. Recall, that this assumption was made to
ensure that rotation curves are asymptotically flat at large radii.

\begin{figure*}
  \includegraphics[width=0.45\textwidth]{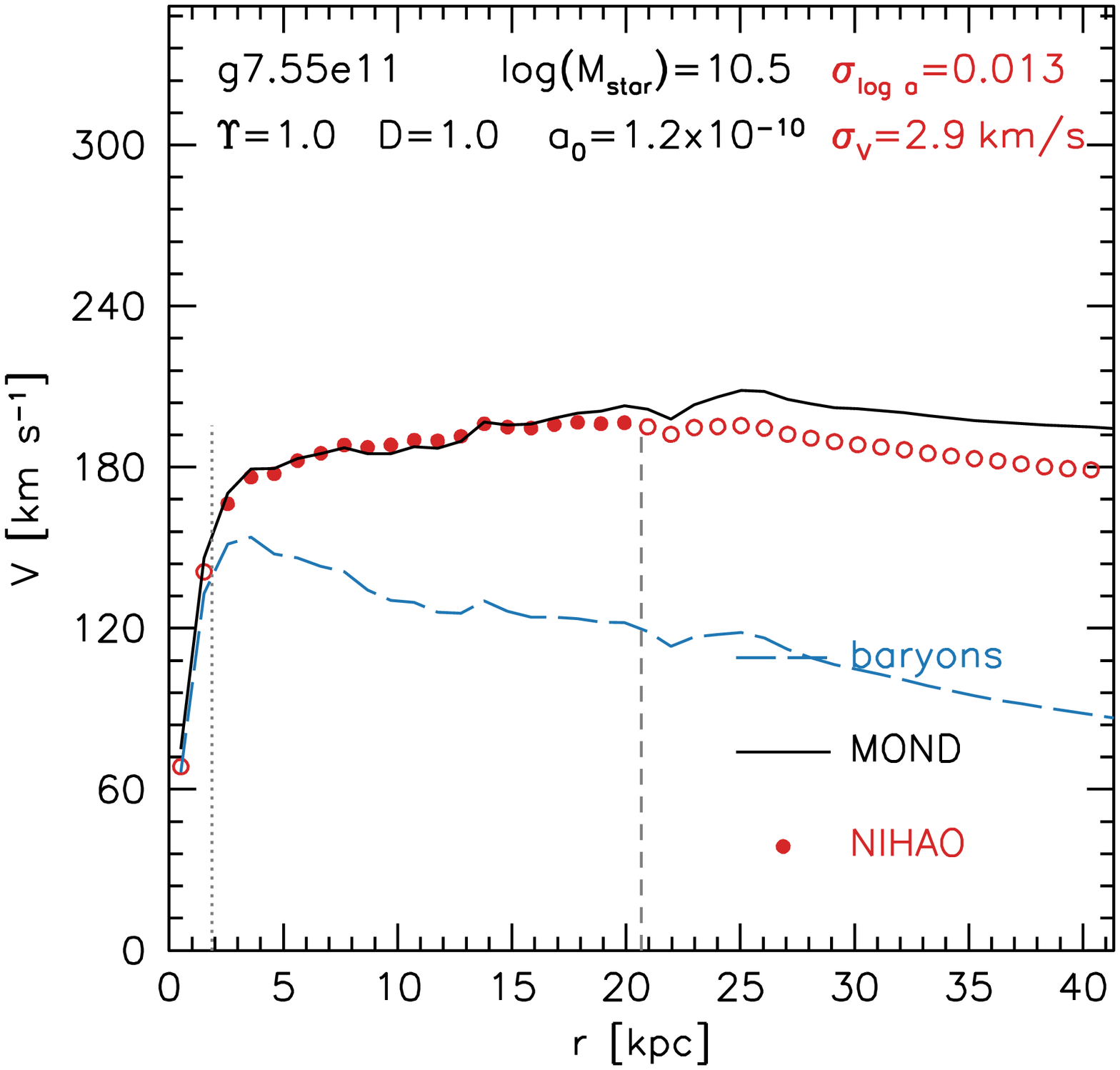}
  \includegraphics[width=0.45\textwidth]{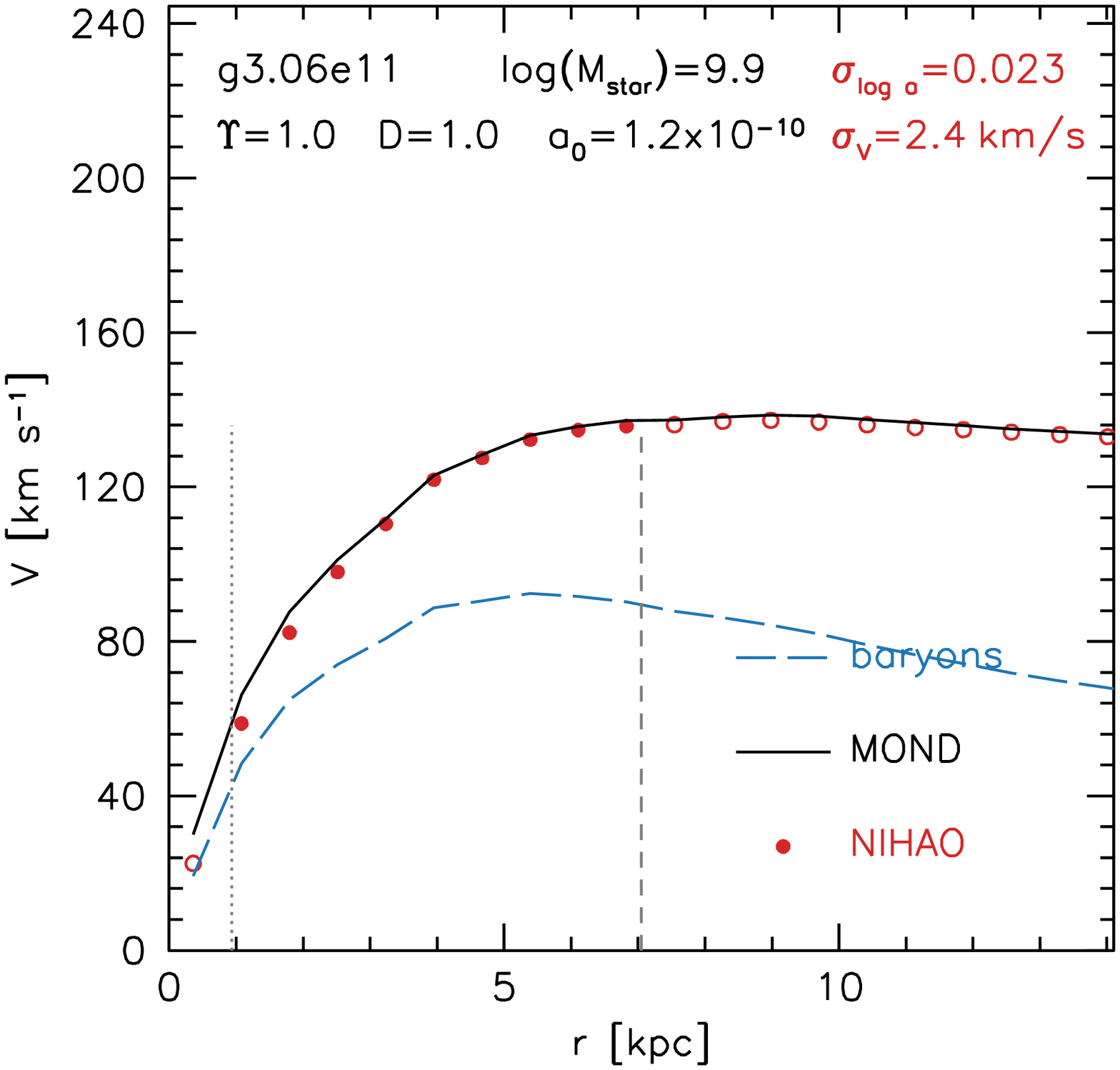}
  \includegraphics[width=0.45\textwidth]{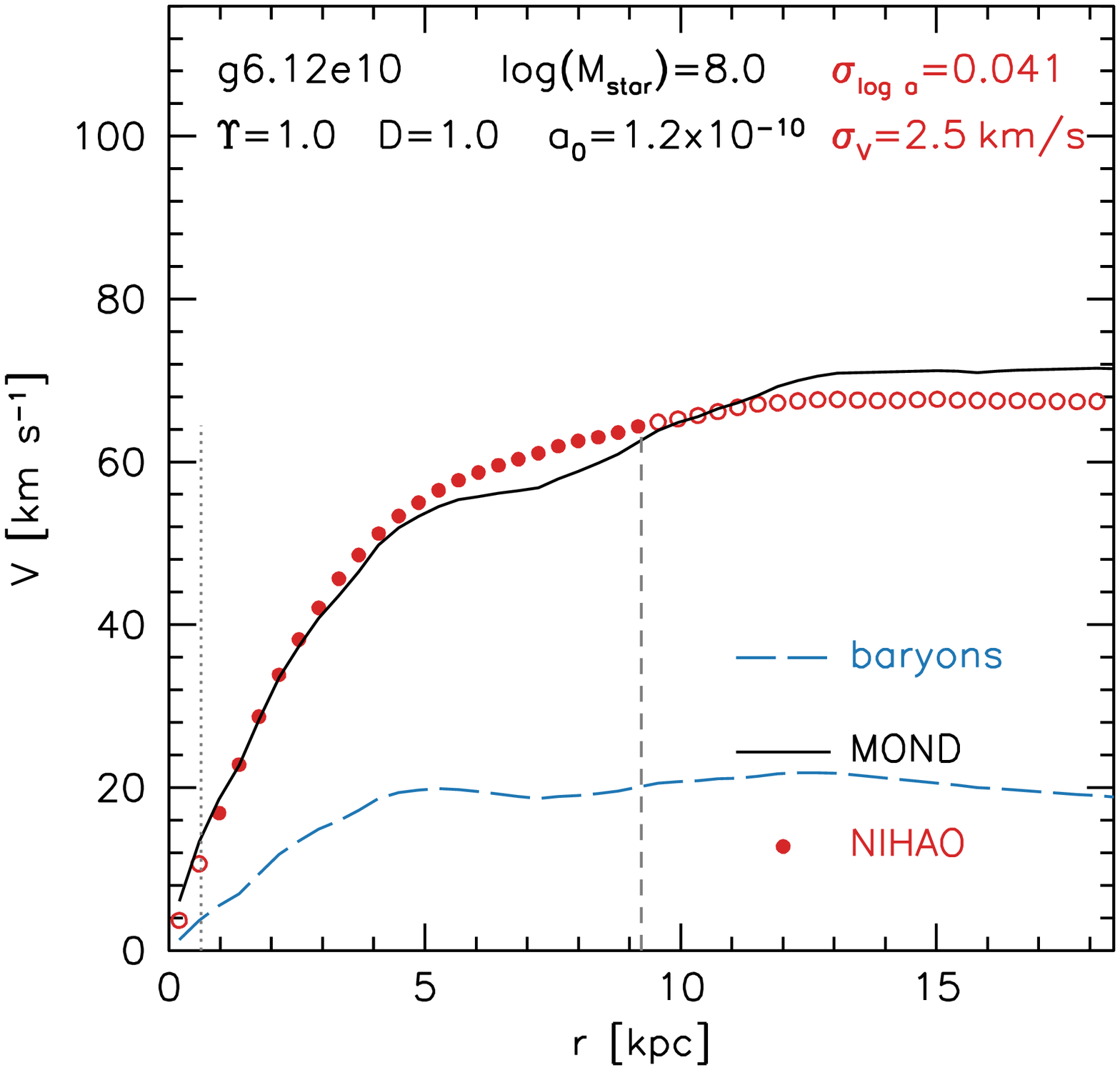}
  \includegraphics[width=0.45\textwidth]{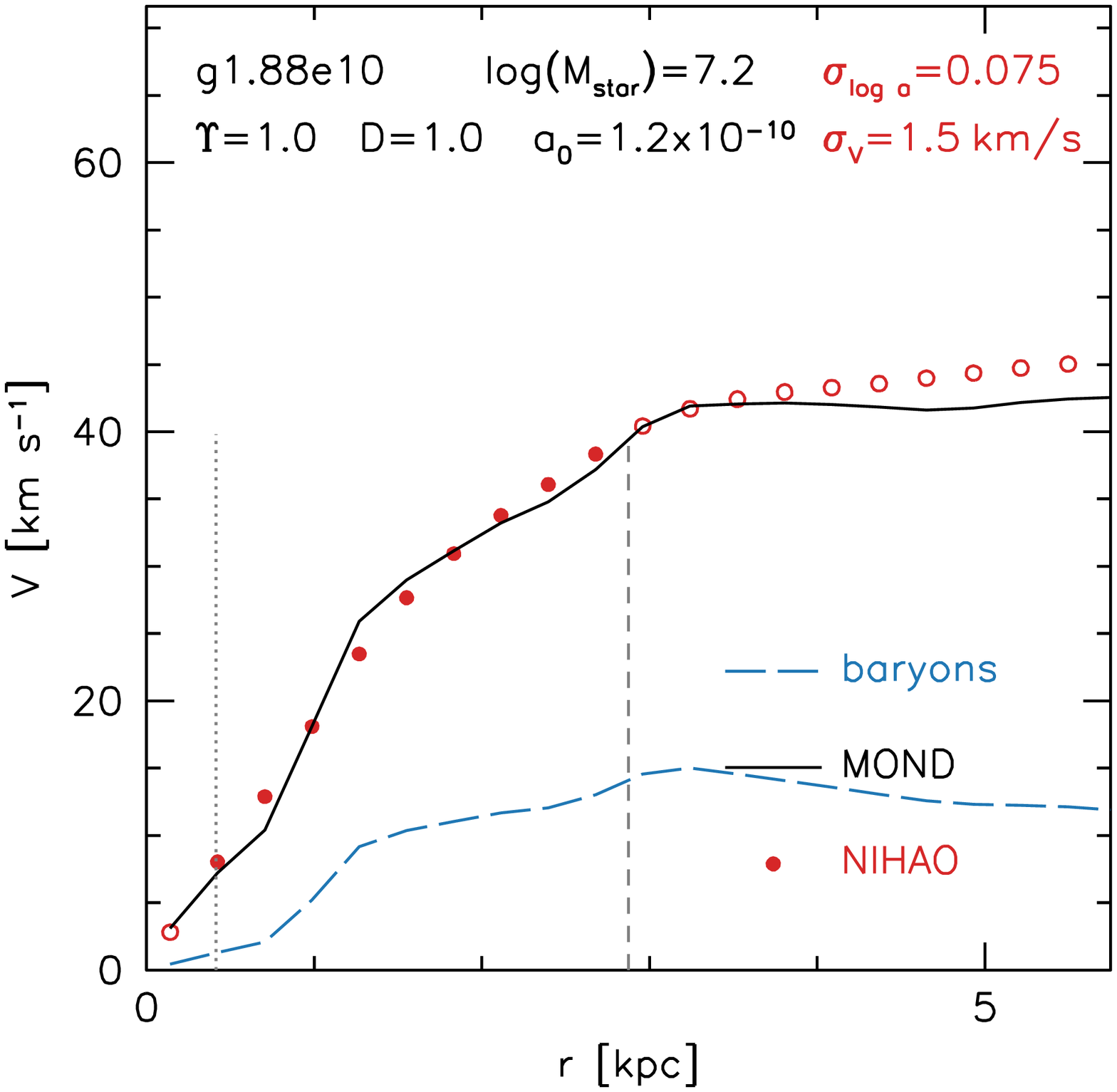}
  \caption{Examples of NIHAO circular velocity curves (red circles)
    where the MOND prediction (black lines) works well. The vertical
    grey lines show twice the dark matter softening length (dotted)
    and the HI radius (dashed). The standard deviation of the 
      velocity residuals, $\sigma_V$, and  acceleration residuals,
    $\sigma_{\rm log a}$, is computed for data points between these
    two lines and is given in the top right corner of each
    panel. These galaxies span over three orders of magnitude in
    stellar masses.}
\label{fig:rotcurve_mond_good}
\end{figure*}

\subsection{Other radial acceleration relations}

Observationally the two axes of the RAR are independent, however in
simulations the two axes are correlated  when the dark matter fraction
is low, because $a\equiv\abar+\adm$.  Fig.~\ref{fig:adm_abar} shows an
alternative way to look at the RAR: the dark matter acceleration vs
the baryon acceleration. This is exactly the same data as in
Fig.~\ref{fig:aabar}, but with independent axes we clearly see that
NIHAO and MOND diverge at high baryon accelerations.  The scatter in
$a_{\rm dm}|\abar$ is also larger than that of $a|\abar$: 0.122 vs
0.079. Observationally it is not advised to calculate the dark matter
acceleration, as measurement errors can result in negative
values. However, it would be possible to determine the scatter in
$a_{\rm dm}$ by forward modelling the observed RAR.

\begin{figure*}
  \includegraphics[width=0.45\textwidth]{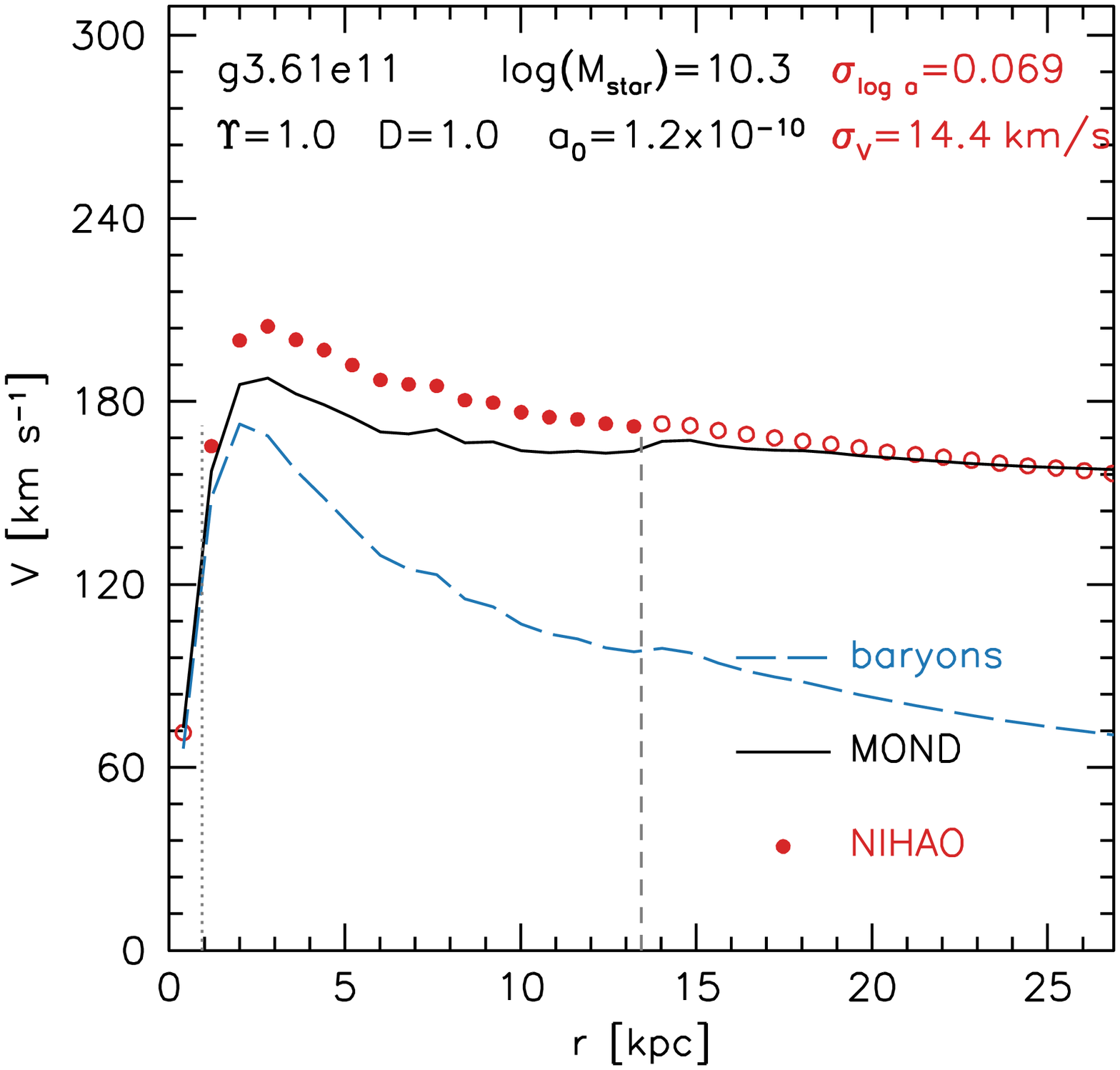}
  \includegraphics[width=0.45\textwidth]{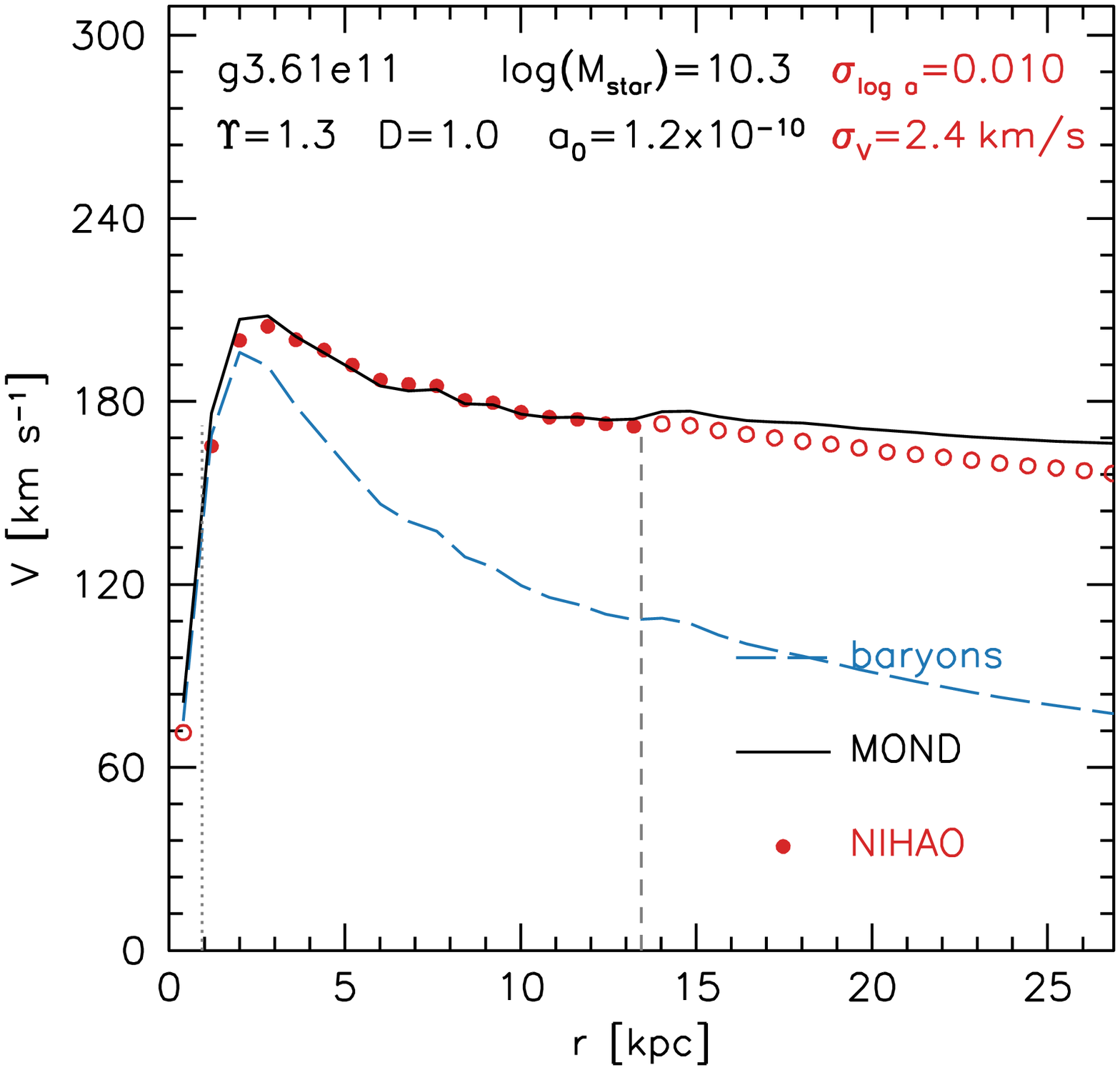}
  \includegraphics[width=0.45\textwidth]{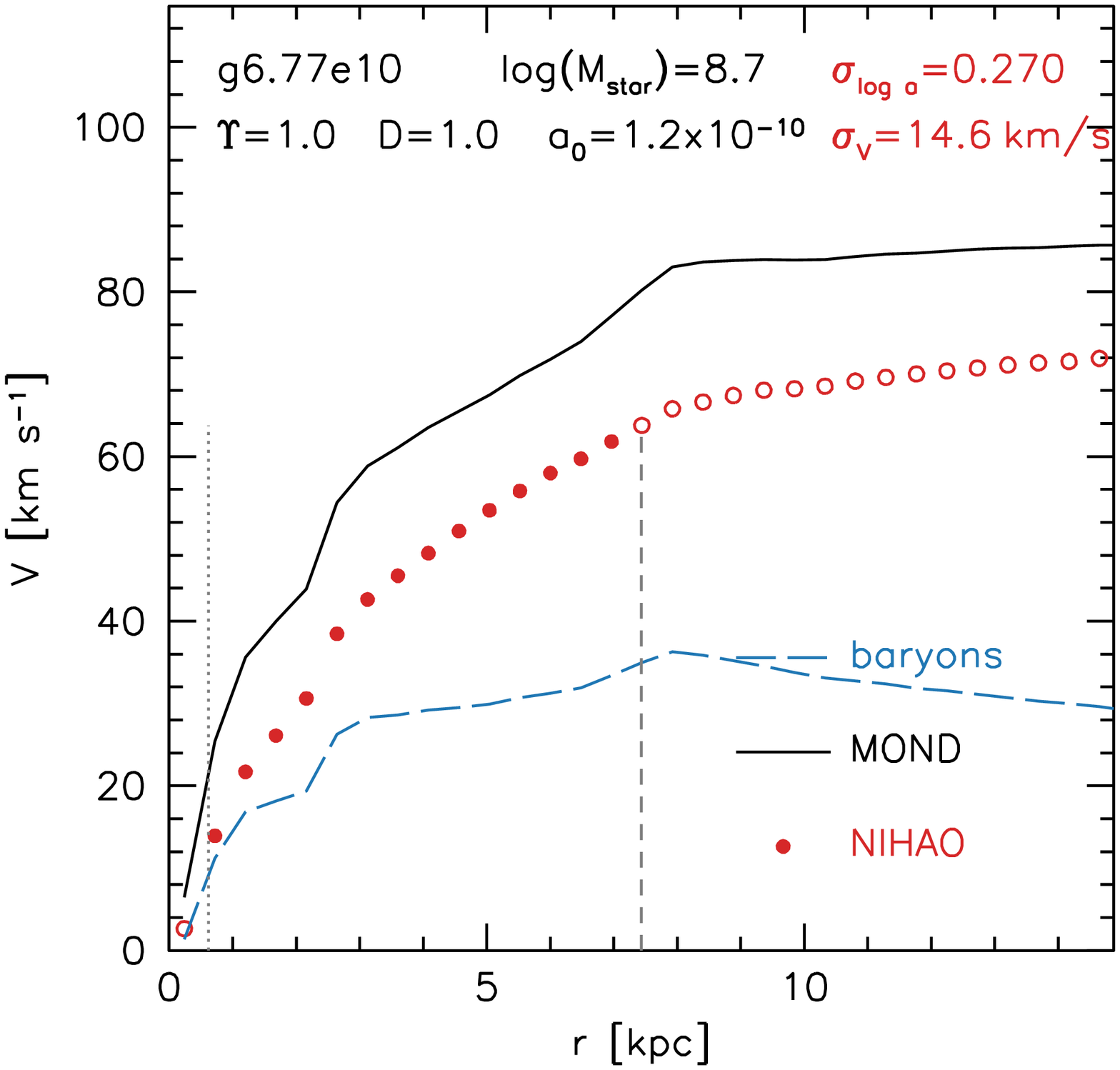}
  \includegraphics[width=0.45\textwidth]{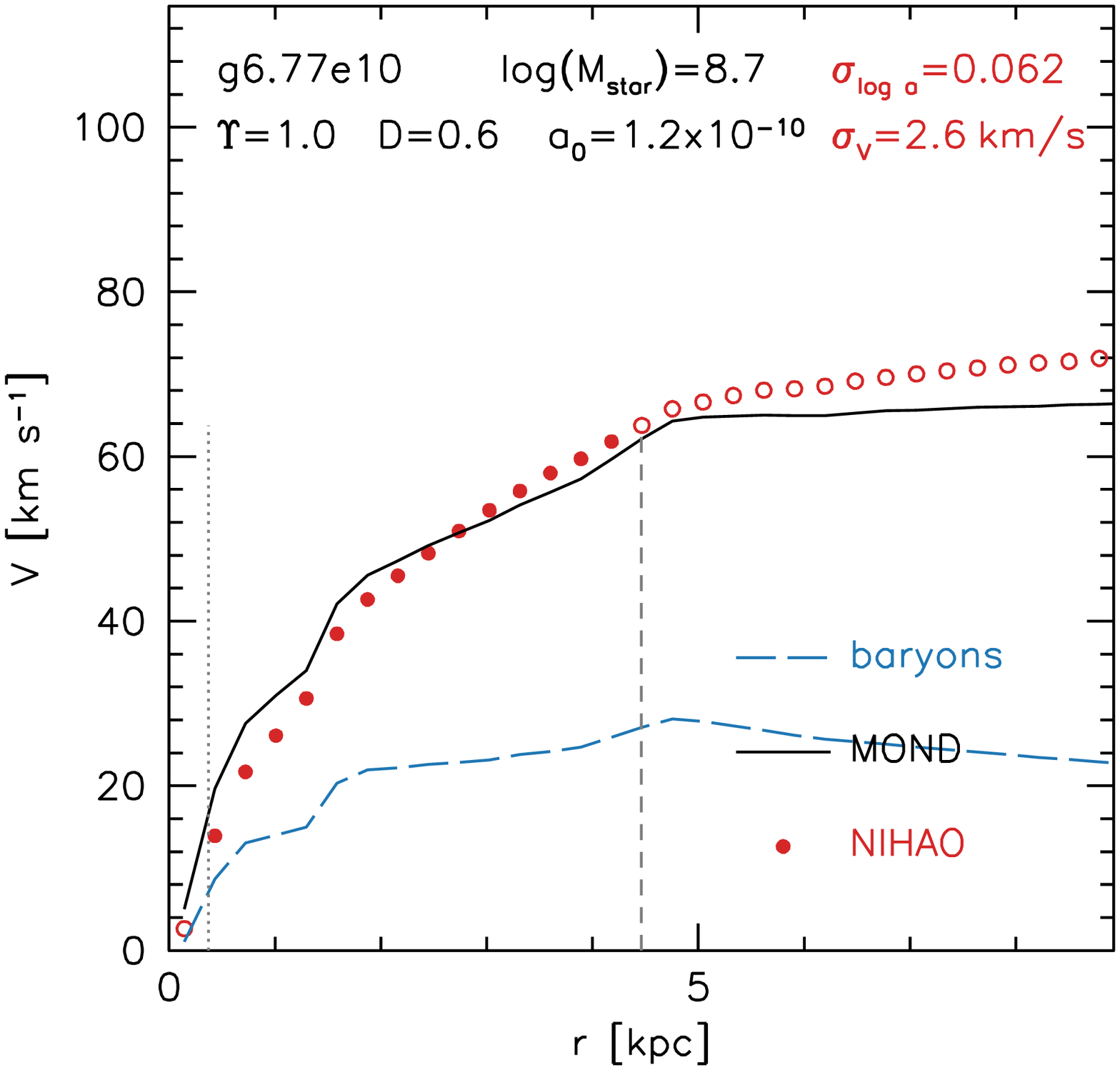}
\caption{Examples of NIHAO circular velocity curves where MOND
  fails. Top left: MOND under predicts the velocity. An excellent fit
  can be obtained by increasing the stellar mass by 30\% (top right),
  mimicking the observational uncertainty in the stellar mass-to-light
  ratio. Bottom left: MOND over predicts the velocity. A good fit can
  be obtained by reducing the distance by 40\% (bottom right). }
\label{fig:rotcurve_mond_bad}
\end{figure*}

\begin{figure}
  \includegraphics[width=0.45\textwidth]{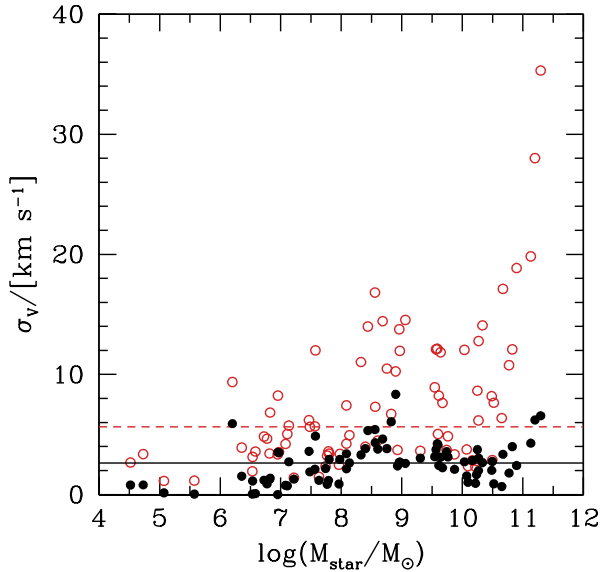}
\caption{Residuals of MOND fits to NIHAO circular velocity
  curves. Fiducial fits (red open circles) which have no free
  parameters, result in a median scatter of just 6 km/s. When the
  stellar mass and ``distance'' are fitted for (black filled circles)
  residuals are reduced even further to 2.5 km/s.}
\label{fig:scatter_rotcurve}
\end{figure}

\begin{figure*}
  \centering{
  \includegraphics[width=0.4\textwidth]{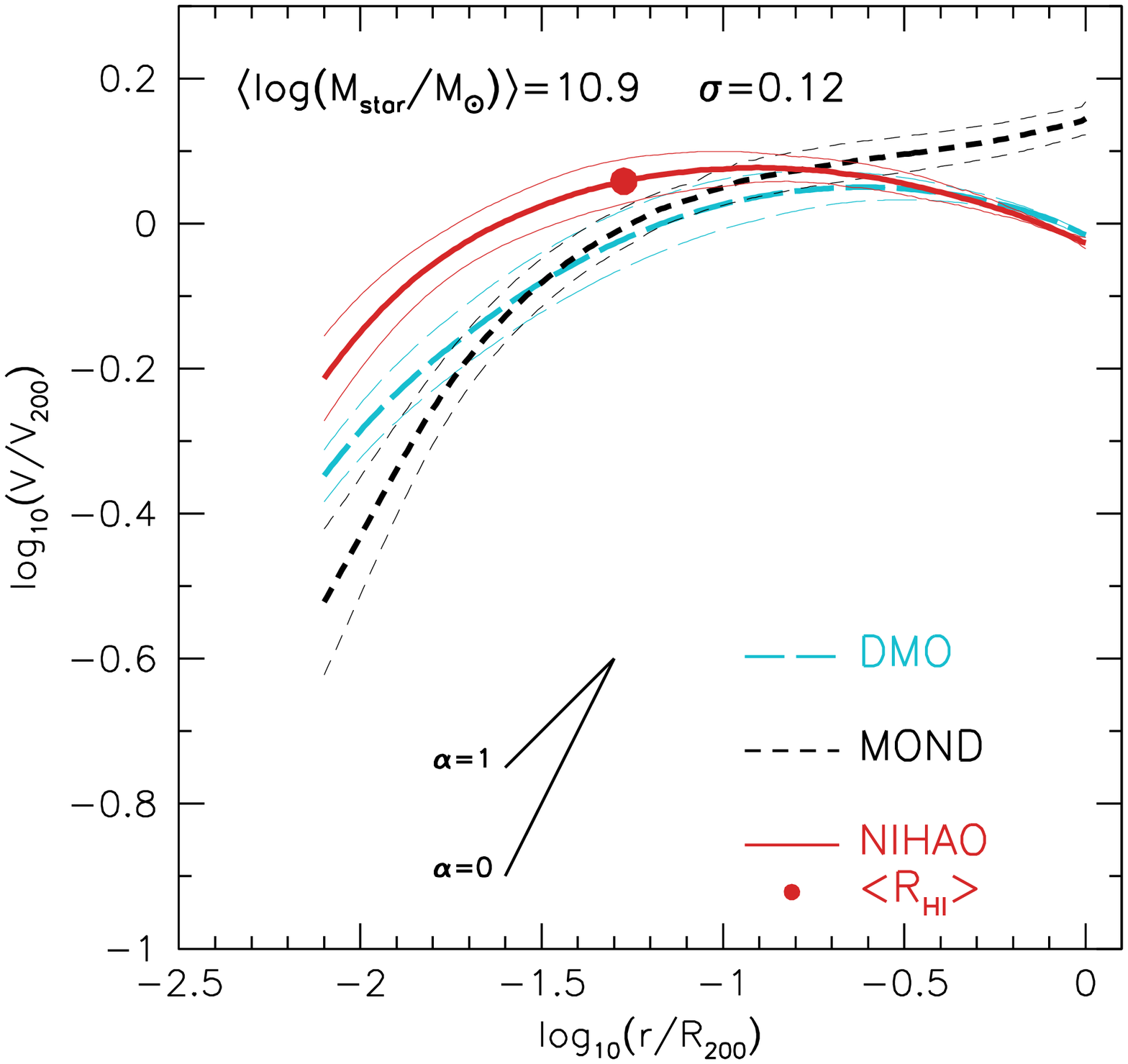}
  \includegraphics[width=0.4\textwidth]{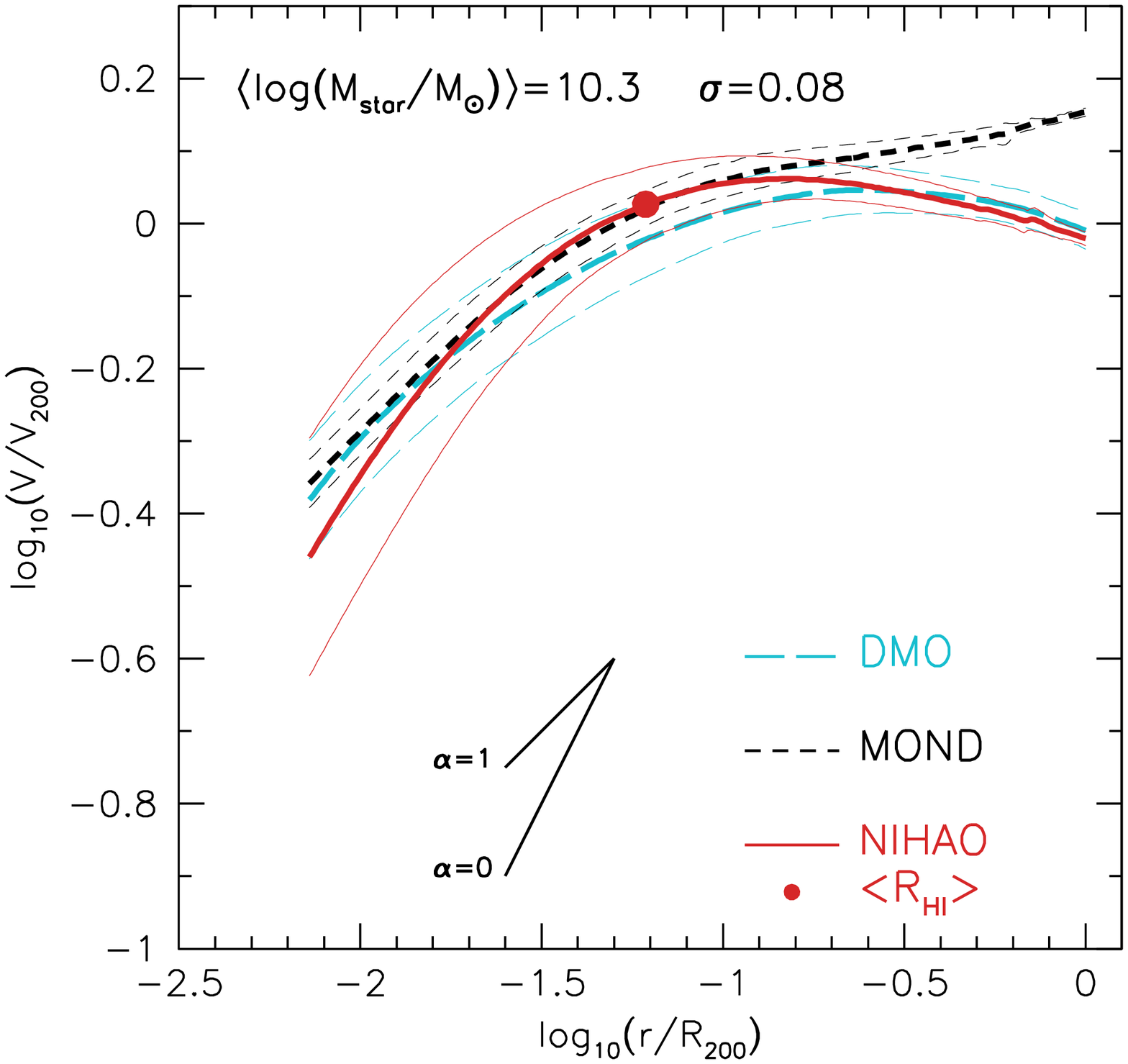}
  \includegraphics[width=0.4\textwidth]{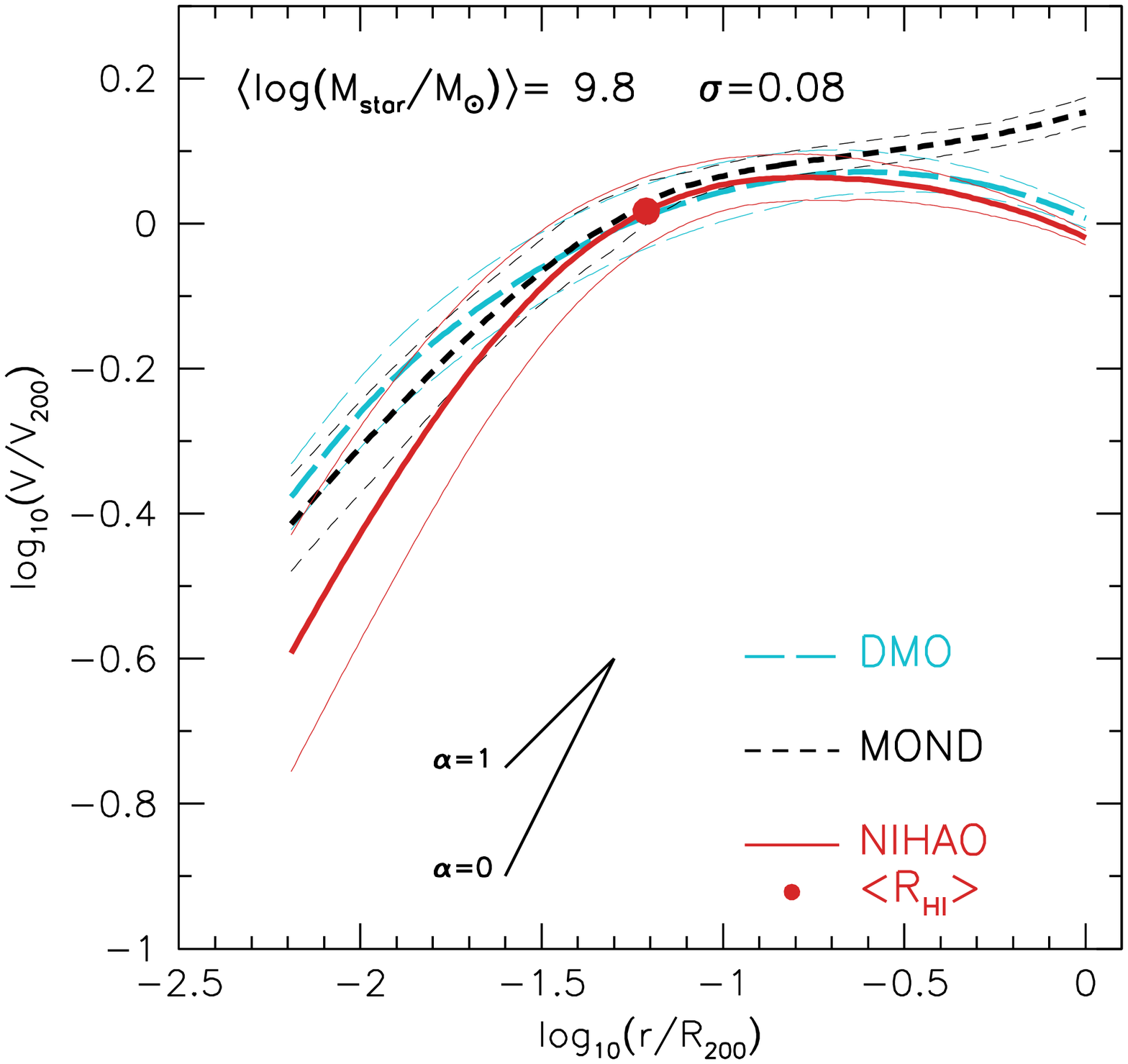}
  \includegraphics[width=0.4\textwidth]{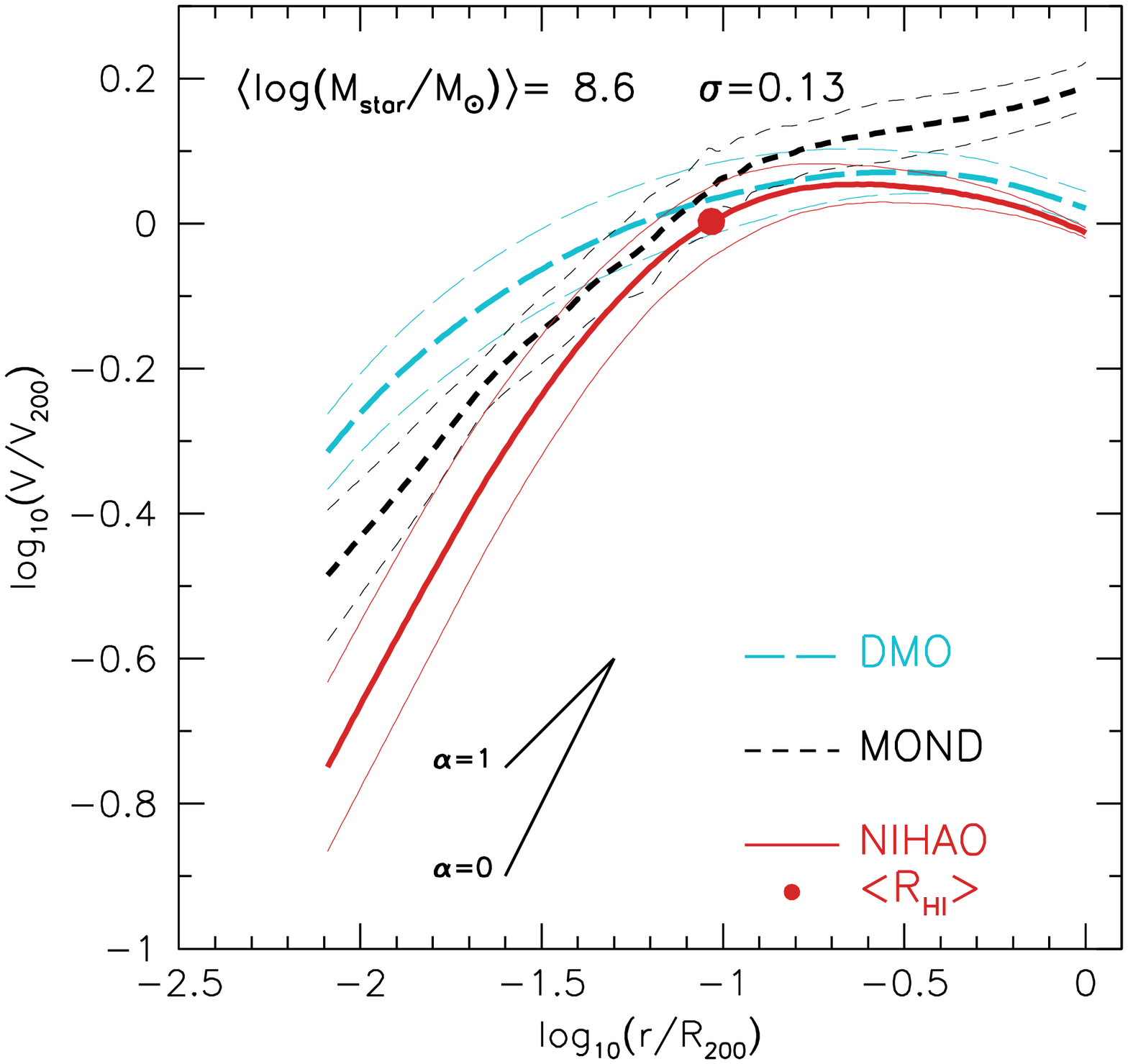}
  \includegraphics[width=0.4\textwidth]{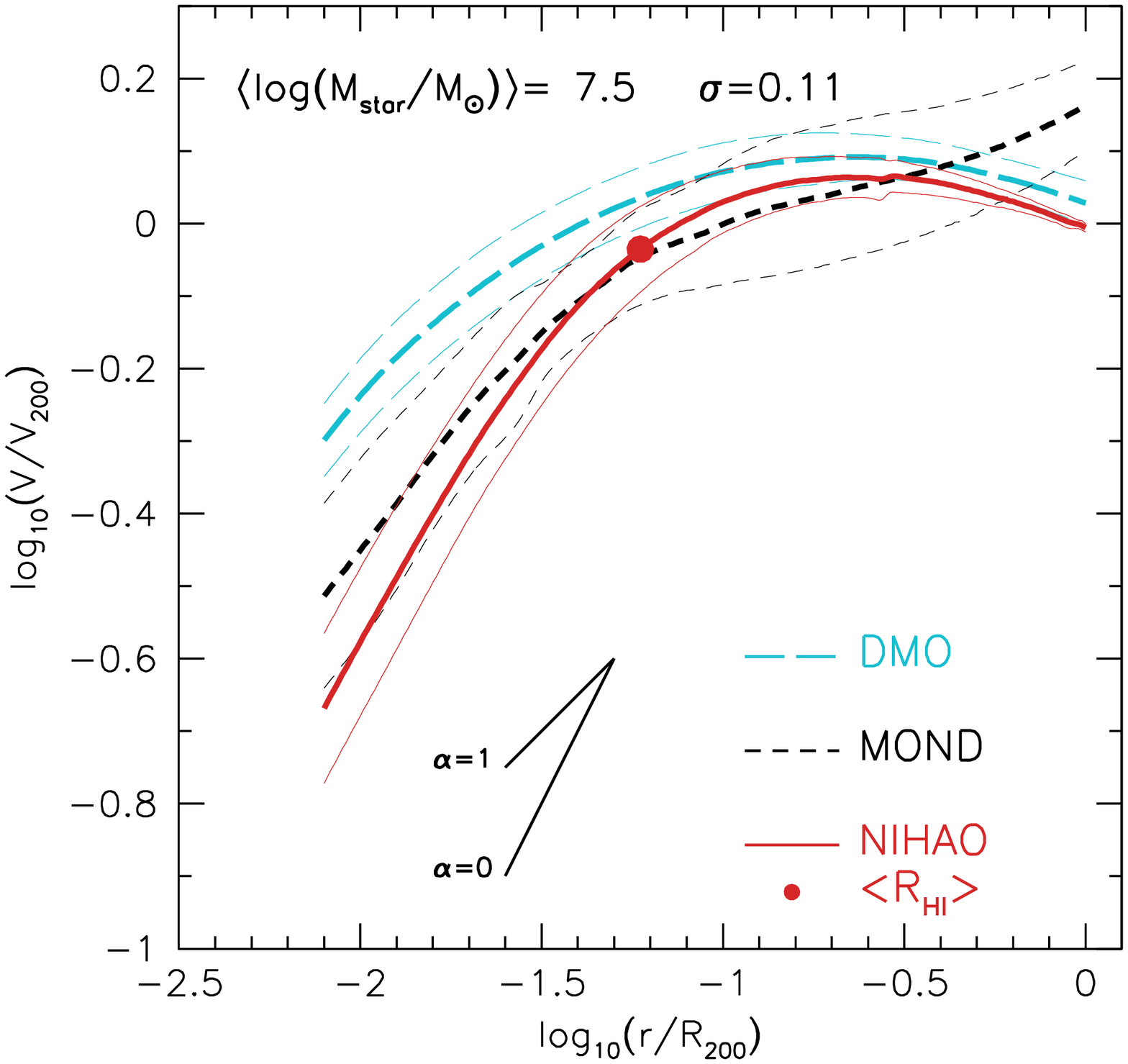}
  \includegraphics[width=0.4\textwidth]{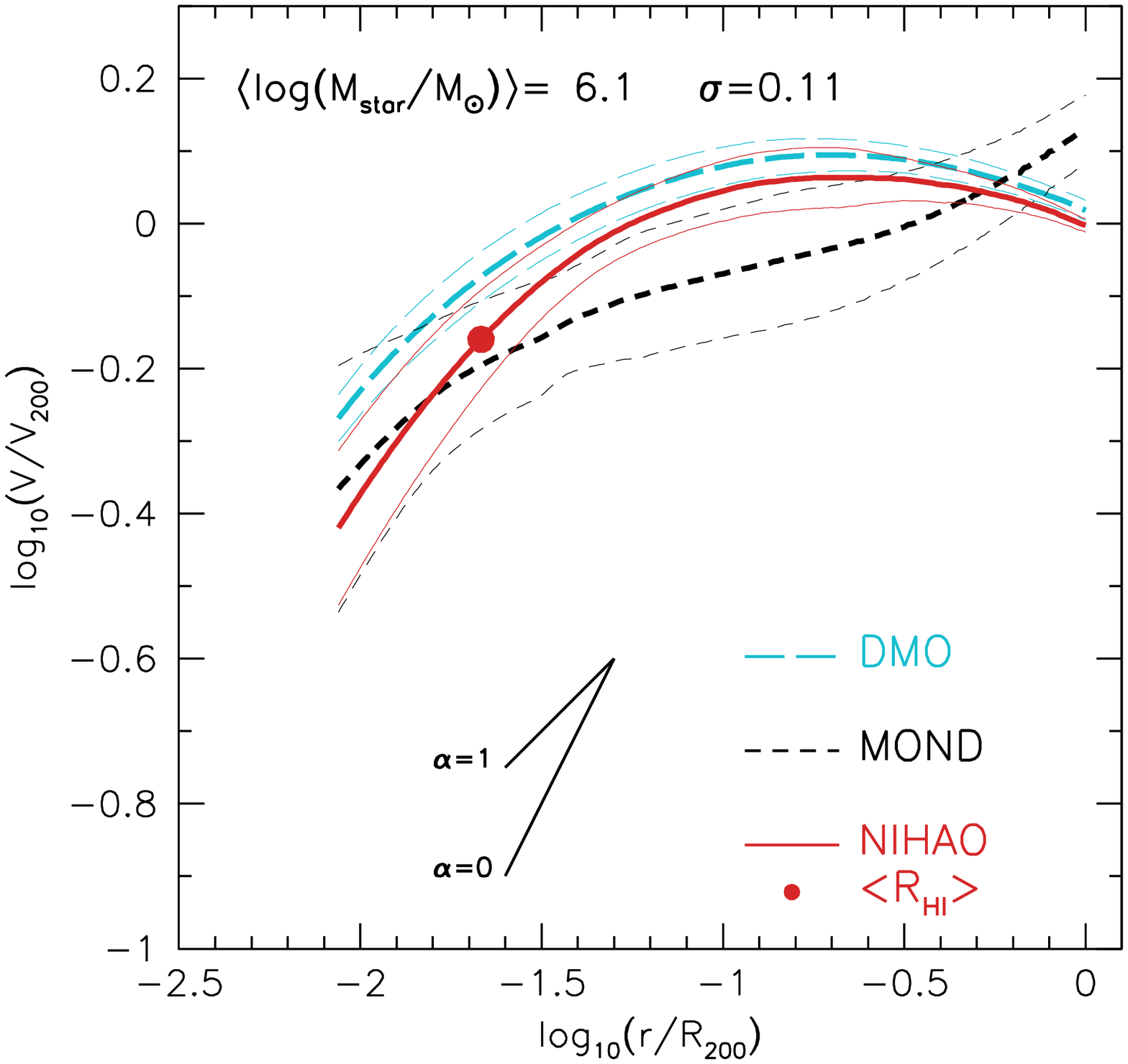}
  }  
  \caption{Dark matter circular velocity profiles from simulations.
    Dark matter only simulation (long-dashed cyan), hydrodynamical
    (red solid), and MOND prediction using the baryon circular
    velocity from the hydrodynamical simulation (short-dashed
    black). Lines are plotted between twice the dark matter softening
    to the virial radius. The red dot shows the average \hi\, radius
    (enclosing 90\% of the \hi\, mass).  Each panel shows galaxies
    with a different range of stellar mass, the average stellar mass
    $\langle\log_{10}(\Mstar/\Msun)\rangle$ is indicated in the top
    left of each panel. The standard deviation between the NIHAO and
    MOND dark matter velocity profiles is given by $\sigma$. For
    reference we show lines with density slopes $\alpha=0$ and
    $\alpha=1$, where $\rho\propto r^{-\alpha}$.  One can see
    systematic deviations between MOND and hydro simulations.}
\label{fig:velocity_nihao_dmo_mond}
\end{figure*}

\section{Origin of RAR scatter in LCDM}
Having established that the NIHAO simulations reproduce the slopes,
normalization,  and small scatter in the RAR, we now ask: where does
the small scatter come from in our simulations?

Since the RAR is a prescription for how the total acceleration is
related to the acceleration due to baryons, we can apply the RAR to
our simulated baryon circular velocity profiles (blue dashed lines in
Fig.~\ref{fig:rotcurve_mond_good}). The baryon circular velocity
profiles are the (quadratic) sum of the circular velocity due to the
stars and gas, calculated from the gravitational potential in the
plane of the disk.

Fig.~\ref{fig:rotcurve_mond_good} shows examples of four simulations
for which the MOND prescription (fitted to the SPARC data) accurately
predicts the circular velocity profile of the simulation. The standard
deviation of $V$ and $\log a$ is given in the top right corner. The
agreement,  to within $3\kms$, is remarkable! We stress that  there
are no free parameters.  The stellar masses of these examples range
over more than three orders of magnitude from $10^{7.2}$ to
$10^{10.5}\Msun$.

However, we have chosen these 4 examples out of a sample of 89
examples.  Fig.~\ref{fig:rotcurve_mond_bad} shows examples of two
simulations for which the MOND prescription clearly fails, one where
MOND under-predicts the velocity (g3.61e11, upper left) and another
where MOND over-predicts (g6.77e10, lower left). For g3.61e11 an
excellent fit can be obtained simply by rescaling the stellar mass
profile by a factor of 1.3 (i.e., 0.11 dex), which is a typical
uncertainty on an observed stellar mass-to-light ratio. For g6.77e10
changing the stellar mass normalization has little impact because the
baryons are dominated by gas. In this case a good fit can be obtained
by reducing the distance by 40\%. This would be a $2\sigma$
uncertainty in distances. One can also change the acceleration scale,
$a_0$, to obtain a similar effect.

Fig.~\ref{fig:scatter_rotcurve} shows the standard deviation of MOND
fits to NIHAO circular velocity profiles.  The fiducial (no free
parameter) fit is shown with red open circles. The median error is
just $6 \kms$.  If we allow the stellar mass and distance to vary, we
can obtain even better fits (black circles), with a median error of
just $2.5 \kms$. This analysis is of relevance to the study of
\citet{Li18} who are able to fit the SPARC sample with a single RAR,
allowing for variation in $M/L$ of the disk and bulge, distance and
inclination.  At first this seems an impressive result for
MOND. However, our similar analysis of NIHAO galaxies shows that MOND
fits have too much freedom, and can fit galaxy velocity profiles that
they should not be able to.

\begin{figure*}
  \includegraphics[width=0.45\textwidth]{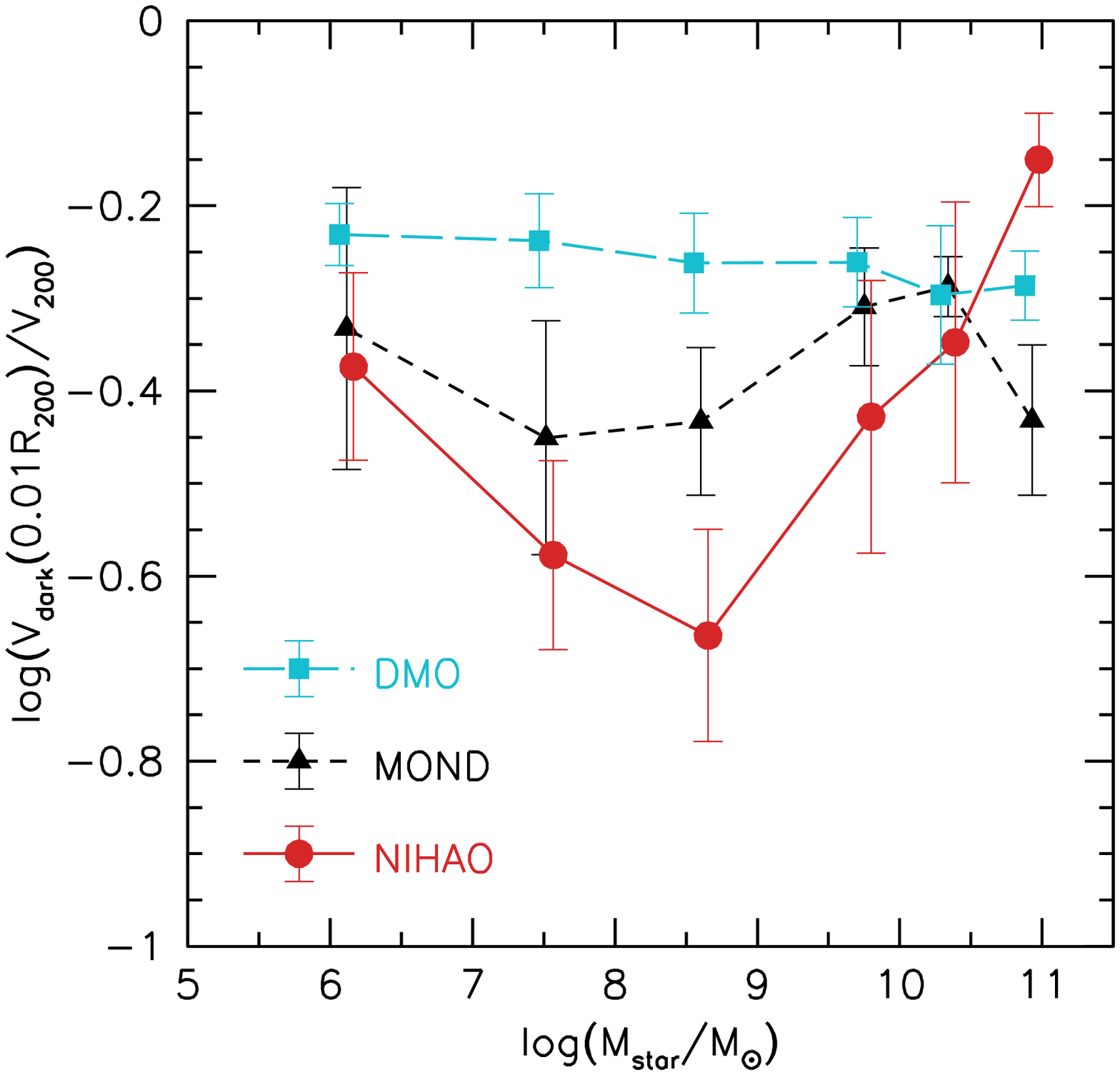}
  \includegraphics[width=0.45\textwidth]{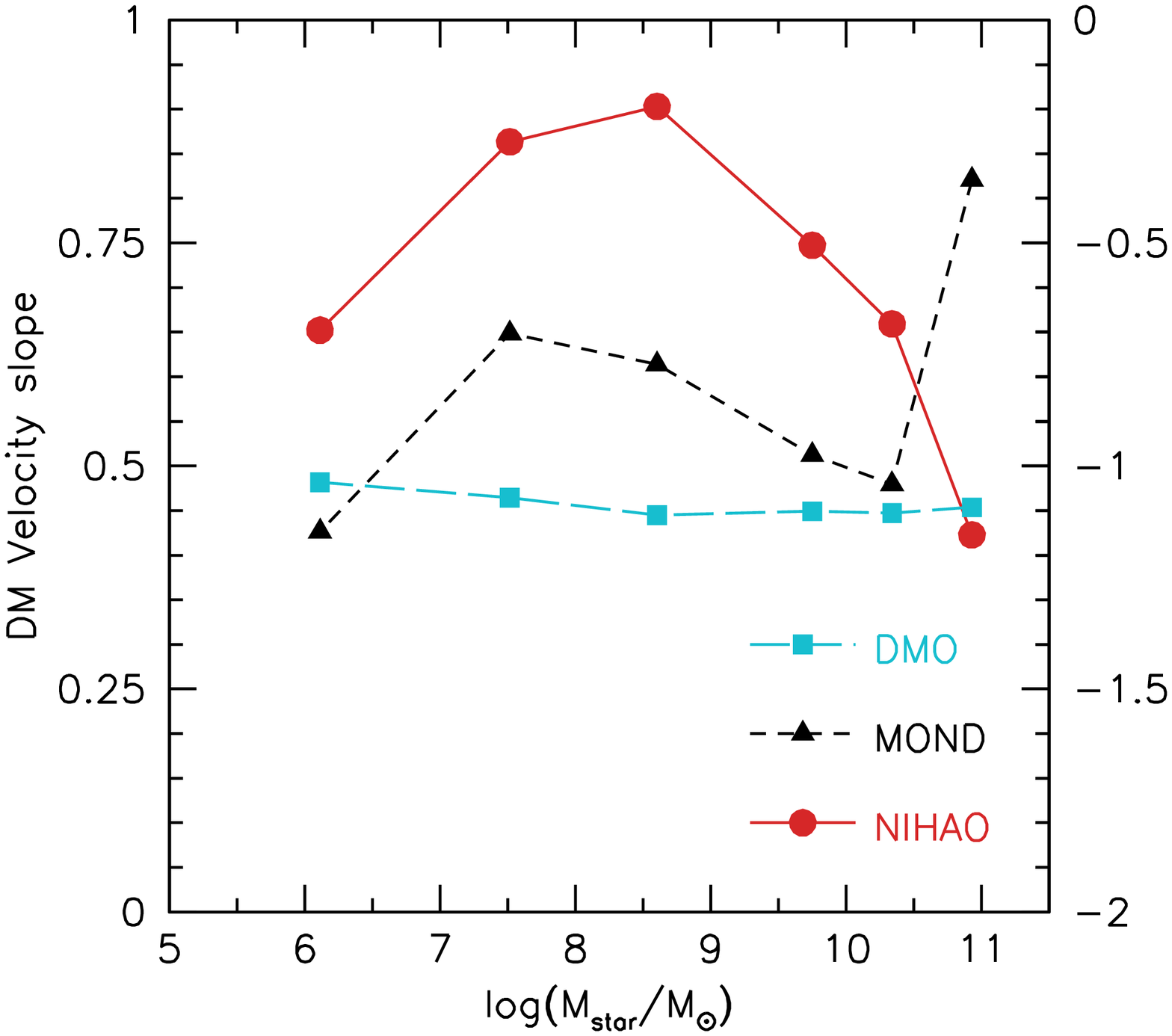}
\caption{Dark matter circular velocity at 1\% of the virial radius
  (left) and dark matter circular velocity slope between 1-2\% of the
  virial radius (right) for NIHAO hydro (red solid), DMO (cyan
  long-dashed), and MOND (black short-dashed). The right axis of the
  right panel shows the enclosed DM density slope.  Stellar mass bins
  are the six panels shown in
  Fig.~\ref{fig:velocity_nihao_dmo_mond}. DMO simulations have a
  velocity and velocity slope independent of the stellar mass. NIHAO
  hydro simulations have contraction at the highest stellar masses,
  and maximum expansion for stellar masses of $\Mstar \sim
  10^9\Msun$. MOND haloes are typically intermediate between DMO and
  Hydro, except for the highest stellar mass, where MOND predicts
  expansion to a core while NIHAO predicts a contracted cusp.}
\label{fig:gamma_dm}
\end{figure*}

These figures also highlight some of the issues with measuring the
scatter in the RAR. Because the scatter is measured in logarithmic
space, galaxies with flat rotation curves tend to have smaller scatter
than those with rising rotation curves where errors at small
velocities get magnified.  The other issue is spatial sampling. Here
we sample galaxies uniformly in units of the virial radius, but we
only include points that are observable (i.e., within the HI
radius). Thus for two galaxies that live in similar mass haloes but
have different HI sizes, the larger galaxy will have more points in
the RAR. This implies that galaxies effectively have different weights
in the RAR. Given the current large uncertainties in the intrinsic
scatter this is a detail, but in the future it will be necessary to
sample the rotation curves in observations and theory in a
consistent way \citep[see e.g.,][]{Desmond17}.

\subsection{Dark matter profiles}
Since the RAR is a prescription of predicting the total acceleration
given the baryon acceleration, it is thus a prescription for
predicting the mass discrepancy or in other words the dark matter
acceleration. When the RAR works well in NIHAO this means that RAR is
accurately predicting the dark matter profile.  By construction, for
the galaxies where the RAR under-predicts the circular velocity, the
RAR under-predicts the dark matter halo, and for the galaxies where
the RAR over-predicts the circular velocity, RAR over-predicts the
dark matter halo. It is then interesting to compare the dark matter
profiles from the RAR with those from the NIHAO and DMO simulations.
This comparison is made in Fig.~\ref{fig:velocity_nihao_dmo_mond}.

Each panel shows a different stellar mass range (corresponding to
those used in Figs.~\ref{fig:aabar2} - \ref{fig:aabar4}), except we
have split the highest mass bin into two to make clearer the different
dark matter profiles at the highest masses we simulate.  The mean
$\log(\Mstar)$ is indicated in the top left of each panel. The thick
lines show mean (of $\log V$) dark matter circular velocity profiles
for NIHAO (red solid), DMO (cyan long dashed) and NIHAO MOND (black
short-dashed). The thin lines show the standard deviation (of $\log
V$).

One of the striking differences between \LCDM (both DMO and NIHAO) and
MOND is that  at large radii (i.e., close to the virial radius) MOND
always over-predicts the circular velocity. This is because of the
assumption that $a^2\propto a_{\rm bar}$ at low accelerations. If all
the baryons were near the center of the halo, the velocity profile
would indeed become flat (which would not be as bad, but still not in
agreement with the declining velocity profiles of the simulations),
but it rises at large radii due to the gas in the halo.  Recall that
in our comparison to SPARC data we only considered radii that can be
traced with atomic hydrogen gas, which typically only extends up to
5-10\% of the virial radius (see the red circles in
Fig.~\ref{fig:velocity_nihao_dmo_mond} which show the average \hi\,
radius). Restricting to the observable radii, the MOND dark matter
profiles typically fall between the DMO and NIHAO lines.

At stellar masses between $10^7\lta \Mstar \lta 10^{10} \Msun$ the
NIHAO simulations have resulted in halo expansion to a constant
density (compare with the reference lines for $\rho\propto
r^{-\alpha}$ with $\alpha=0$ and $\alpha=1$), see also
\citet{Tollet16}.  MOND also tends to predicts expanded haloes, but
not as much as in NIHAO.

For the highest masses ($\Mstar=10^{10.9}\Msun$, top left panel)
the NIHAO simulations have contracted dark haloes, while MOND predicts
expansion. These are examples similar to that shown in the top left of
Fig.~\ref{fig:rotcurve_mond_bad}, where this deficit in dark matter
can be compensated for with increased stellar mass.  In spite of these
differences, because the centers of these galaxies are dominated by
baryons, they are only offset by a small amount from the SPARC
RAR. The average rms in $a$ of these 8 galaxies is 0.056 dex, making
them typical galaxies.  Determining the dark matter profiles
observationally, and thus distinguishing between the predictions from
MOND and NIHAO requires an accurate determination of the stellar
mass-to-light ratio, which is a challenge.

A summary of the density profiles is shown in Fig.~\ref{fig:gamma_dm}.
This shows the dark matter circular velocity at 1\% of the virial
radius vs stellar mass (left panel) and the slope of the dark matter
circular velocity between 1 and 2\% of the virial radius
[$\gamma_{V}=\log_{10}(V_{0.02}/V_{0.01})/\log_{10}(2)$] vs stellar
mass (right panel).  Both plots show the same qualitative trends.  DMO
simulations (cyan) have structure that is almost scale free, with
velocity $\log_{10}[V_{\rm dark}(0.01 R_{200})/V_{200}]\simeq -0.25$
and velocity slope $\gamma_{\rm V}\simeq 0.5$.  Note that we choose to
show velocity slope rather than (local) density slope because the
latter involves an extra derivative of the velocity profile, and so is
noisier in both observations and simulations. The circular velocity
profile, or equivalently the cumulative mass profile, or cumulative
enclosed mass density profile, also has advantages in terms of
analytic models \citep{Dekel17}.  The velocity slope can be easily
converted into an enclosed mass density slope using $\gamma_{\rho} =
2\gamma_V - 2$. So for example a constant density core,
$\gamma_{\rho}=0$ corresponds to a velocity slope of $\gamma_V=1$, a
NFW density slope of $\gamma_{\rho}=-1$ corresponds to a velocity
slope of $\gamma_V=0.5$, and an isothermal density slope of
$\gamma_{\rho}=-2$ corresponds to a velocity slope of $\gamma_V=0$.

In NIHAO hydro simulations (red points and lines) the halo response is
strongly mass dependent  \citep[See also][]{Tollet16, Dutton16}. At
low stellar masses $\Mstar \lta 10^6\Msun$ there is  very little halo
response. As the stellar mass increases the halo
expands (i.e., lower velocity and higher velocity slope), reaching a
maximum expansion at stellar masses of $\Mstar \sim 10^9\Msun$, where the
velocities are $\simeq 0.4$ dex lower and the velocity slopes are 0.4
higher than DMO and close to a constant density core ($\gamma_{V}=1$).  By
a mass of $\Mstar \sim 10^{10.5}\Msun$ the haloes are on average unchanged at
small radii, and at the highest stellar masses $\Mstar \sim 10^{11}\Msun$ the
haloes contract at small radii.   The MOND dark matter haloes (black
points and lines) typically show behavior intermediate between NIHAO
and DMO (i.e., the haloes are not as cuspy and DMO, and not as
expanded as NIHAO). The exception is at the highest masses, where MOND
haloes have lower velocities and higher slopes, the opposite of NIHAO,
which has higher velocities and lower slopes.

\section{Summary}
\label{sec:sum}

We use 89 cosmological galaxy formation simulations from the NIHAO
project \citep{Wang15} to study the origin of the radial acceleration
relation (RAR) in a \LCDM universe.  The unique combination of halo
mass range, resolution, and number of haloes allows us to compare with
observations of the RAR from the SPARC survey \citep{McGaugh16}.  The
NIHAO galaxies have stellar masses spanning $10^{4.5}\lta \Mstar \lta
10^{11.3}\Msun$.

We summarize our results as follows:

\begin{itemize}
\item The RAR exists in the NIHAO simulations with a similar slope and
  normalization as observationally determined by SPARC
  (Fig.~\ref{fig:aabar}).
  
\item The RAR in NIHAO has a scatter of 0.079 dex. This is consistent
  with estimates of the intrinsic scatter from observations ($0.0 \lta
  \sigma_{\rm int}\lta 0.09$) (Fig.~\ref{fig:rar_scatter}).

\item The scatter in the RAR depends on the stellar mass of the
  galaxy, with lower scatter in higher mass galaxies in both
  observations and simulations (Fig.~\ref{fig:rar_scatter_mass}).  For
  high mass galaxies ($\Mstar \gta 10^{9.3}\Msun$) the intrinsic
  scatter ($\sigma_{\rm int}\lta 0.05$) is consistent with both MOND
  and NIHAO. However, for low mass galaxies ($\Mstar \lta
  10^{9.3}\Msun$) the intrinsic scatter ($0.10\lta\sigma_{\rm int}\lta
  0.15$) is consistent with NIHAO, but inconsistent with MOND.
  
\item The mass dependence of the RAR scatter in simulations is
  partially explained by the correlation between $a\equiv\abar+\adm$
  and $\abar$ at low dark matter fractions.
  
\item In the lowest mass galaxies we simulate ($\Mstar\sim
  10^{6}\Msun$) The RAR deviates significantly from the MOND
  prediction, but is in agreement with the observed RAR from dwarf
  spheroidal galaxies (Fig.~\ref{fig:aabar4}).
  
\item We can use the RAR to accurately predict (to within a few
  $\kms$) the circular velocity profiles of individual NIHAO
  simulations by just using the baryon circular velocity profile as
  input (Fig.~\ref{fig:rotcurve_mond_good}). 

\item  In cases where the fiducial values provide a bad fit, the
  stellar masses can be re-scaled, and/or the distances changed to
  provide excellent fits (Fig.~\ref{fig:rotcurve_mond_bad}). Given
  realistic uncertainties in baryon profiles and galaxy distances,
  MOND has too much flexibility in fitting individual galaxy rotation
  curves. Thus the success of MOND at fitting individual galaxy
  rotation curves is not as impressive as it is often claimed to be
  \citep[e.g.,][]{Sanders02, Li18}.

\item  The RAR predicts dark matter circular velocity profiles that
  are on average similar to that found in \LCDM simulations
  (Fig.~\ref{fig:velocity_nihao_dmo_mond}).

\item  In detail, the RAR generally predicts mild halo expansion at
  small radii (relative to dissipationless CDM), but not as much
  expansion as found in the NIHAO simulations
  (Fig.~\ref{fig:gamma_dm}). This contradicts the claim by
  \citet{Navarro17} that explaining the RAR in CDM does not require
  modifications to the cuspy inner mass profiles of dark haloes.
  
\item The largest differences in the dark matter profiles are in the
  highest mass galaxies we simulate.  For ($\Mstar\sim
  10^{11}\Msun$) NIHAO predicts halo contraction, while the RAR
  predicts expansion to a dark matter core
  (Figs.~\ref{fig:velocity_nihao_dmo_mond} \& \ref{fig:gamma_dm}).
  
\end{itemize}

It may seem paradoxical that the MOND RAR appears to be in the NIHAO
\LCDM galaxy formation simulations. The resolution is both MOND and
NIHAO are approximating the same thing (the observable Universe).

Since the phenomenological basis for MOND is naturally reproduced by
\LCDM galaxy formation simulations, it seems unnecessary to invoke the
simplifying assumptions and radical new force law required by MOND.  A
useful analogy is with solar system. A phenomenological theory might
assume that the orbits of planets are circles, since they are observed
to be roughly circles, and this is the simplest geometry. But we know
that with sufficiently accurate observations the orbits are actually
ellipses, and the circle model is an over-simplification.

The assumptions MOND is based on are only approximately true in
$\LCDM$: 1) At high accelerations the dark matter fraction is low, but
non-zero; 2) Circular velocity profiles are only approximately flat at
the HI radius of a galaxy. At larger radii the profiles decline;  3)
The scatter in the RAR is small, but non-zero, primarily reflecting
non-universality of the dark matter density profiles.  Finally, we
note that if one wishes to go beyond Newtonian dynamics there is more
freedom in the observed RAR than imposed by MOND.

\section*{Acknowledgments} 
 We thank the referee for a prompt report that led to significant
 improvements in the paper. The authors gratefully acknowledge the
 Gauss Centre for Supercomputing e.V. (www.gauss-centre.eu) for
 funding this project by providing computing time on the GCS
 Supercomputer SuperMUC at Leibniz Supercomputing Centre (www.lrz.de).
 Part of this research was also carried out on the High Performance
 Computing resources at New York University Abu Dhabi; on the {\sc
   theo} cluster of the Max-Planck-Institut f\"ur Astronomie; and on
 the {\sc hydra} cluster at the Rechenzentrum in Garching.  We greatly
 appreciate the contributions of all these computing allocations.
AO is funded by the Deutsche Forschungsgemeinschaft (DFG, German Research Foundation) -- MO 2979/1-1.
TB was supported by the Sonderforschungsbereich SFB 881 ``The Milky
Way System'' (subprojects A1 and A2) of the DFG.
The analysis made use of the {\sc pynbody} package \citep{Pontzen13}.





\bsp	
\label{lastpage}
\end{document}